\documentstyle[aaspp4,12pt]{article}
\def\Msun{\hbox{M$_{\odot}$ }}

\def\etal{{\it et al.}}
\begin{document}

\title{The Dynamical Evolution of Dense Rotating Systems\\
Paper I.  Two-Body Relaxation Effects}
\author{John S. Arabadjis and Douglas O. Richstone}
\affil{University of Michigan Department of Astronomy}
\authoraddr{Ann Arbor, MI 48109}

\begin{abstract}

% old abstract
%The evolution of dense rotating systems is studied through the use of N-body
%simulations.  The purpose of this paper is twofold -- to test the programs
%used here by comparing results to previous studies, and to disentangle mass
%segregation effects from other processes which will be treated in a subsequent
%study (Paper II).  The initial configuration for each experiment is an
%isotropic Kuzmin-Kutuzov model.  Hernquist's tree code is utilized to simulate
%the dynamical evolution, with dynamical relaxation produced by the graininess
%of the low-N potential field.  Consistent with previous studies, we find that
%systems whose constituent particles follow the Salpeter initial mass function
%rapidly undergo core collapse through Spitzer's mass segregation instability.
%We find that the lump of high mass stars which condenses at the center wanders
%about the core in Brownian motion.  All rotationally flattened systems show a
%decrease in flattening with time, consistent with Fokker-Planck calculations.
%Angular momentum transport is directed outward; systems containing equal-mass
%stars transport angular momentum in the core more efficiently than do systems
%following a Salpeter spectrum.

%new abstract
This paper, and its companion, investigate the evolution of dense
stellar systems due to the influence of two-body gravitational
encounters, physical collisions and stellar evolution.  Our goal is
the simulation of the densest centers of galaxies, like M32, which
reach stellar densities near $10^6 \Msun/{\rm pc}^3$ and which may
harbor black holes.  These systems have a different Safronov number
(the dimensionless ratio of stellar binding energy to mean stellar
kinetic energy) than globular clusters, substantially increasing the
importance of physical collisions relative to gravitational
encounters.  In this paper, we focus only on the gravitational
encounters.  We demonstrate, first, that our simulations with small N
with a Hernquist tree code yield results basically in accord with years
of effort studying globular clusters.  Second, we investigate
(crudely) core collapse in rotating systems with mass segregation, to
separate out the effects purely due to two-body encounters from those
seen in the more complex second paper.

The initial configuration for each experiment is an isotropic
Kuzmin-Kutuzov model.  Hernquist's tree code is utilized to simulate
the dynamical evolution, with dynamical relaxation produced by the
graininess of the low-N potential field.  Consistent with previous
studies, we find that systems whose constituent particles follow the
Salpeter initial mass function rapidly undergo core collapse through
Spitzer's mass segregation instability.  All rotationally flattened
systems show a decrease in flattening with time, consistent with
Fokker-Planck calculations.

An interesting new result concerns the well-established inability of
simulators to identify a static center in their simulations of
collisional systems.  We find that the lump of high mass stars which
condenses at the center wanders about the core in Brownian motion.  We
see this most clearly in the simulations with a mass function, in
which the lump (possibly abetted by our force softening) lives on as a
discrete entity after its formation.  Even in the equal-mass case a
similar but transient concentration appears and jitters about the
center with similar radii to the unequal-mass case.

\end{abstract}


\section{Introduction}

There are many physical processes which govern the evolution of a
dense stellar system.  Two-body relaxation, stellar evolution and mass
loss, star formation, and stellar collisions all influence the
evolution of the system.  We examine each of these processes through
the use of N-body simulations.  In order to discriminate among effects
caused by these processes we have divided the study into two parts.
This study (Paper I) examines the effects caused by two-body
gravitational scattering, system rotation, and a stellar mass spectrum
-- the processes that do not depend on the finite size of the
interacting particles.  The purpose of this section of the work is
twofold. First, we wish to demonstrate that our implementation of this
technique yields results consistent with the extensive body of work on
globular clusters, since the absence of such consistency would cast
doubt on the more complicated results in paper II.  Of course, we are
also interested in any new effects that come out of this treatment of
rotation, although they turn out to be rather meager.  In Paper II we
investigate the effects of stellar collisions, stellar evolution and
mass loss, star formation, and a central black hole.  Since several of
the processes examined in Paper II either compete with or enhance mass
segregation, it is necessary to examine them separately in order to
determine their relative importance to the evolution of a dense
stellar system.

An examination of these processes may help to explain the structure of the
dwarf elliptical M32, an E2 galaxy containing no dust or disk structure.
Its proximity allows a detailed study of its kinematics.  Hubble Space
Telescope observations indicate a maximum core size of 0.37 pc and a central
density in excess of $4 \times 10^6$ \Msun pc$^{-3}$ (Lauer \etal\ 1992).
For comparison, the globular cluster M13 has corresponding values of
$r_c = 1.8$ pc and $\rho_c = 2.0\times10^3$ \Msun pc$^{-3}$ (Djorgovsky
1993).  The central velocity dispersion of M32 is 126 km s$^{-1}$, and the
rotation speed is 50 km s$^{-1}$ less than 4 pc from the center (van der
Marel \etal\ 1997, 1994b).  Dressler and Richstone (1988) have ruled out
constant M/L models of M32 using a maximum entropy technique; we are left with
two kinds of models: those which contain a central black hole, and those with
a central cluster of either low-mass or degenerate stars (e.g. Goodman and Lee
1989, hereafter GL).  Richstone {\it et al.} (1990) have shown that the
observations are consistent with a central black hole of mass
0.7-3$\times 10^6$ \Msun or a $10^7$ \Msun dark cluster with a core radius
$\leq 0.''4$ (1.3 pc).  Subsequent studies (van der Marel \etal\ 1994a, Qian
\etal\ 1995) showed that a good fit to the data is achieved with models
containing a central black hole of mass of $1.8\times 10^6$ \Msun.  Recently
van der Marel \etal\ (1997) have derived a mass of $3\times 10^6$ \Msun\ for
the black hole.

The time scales for two-body relaxation and physical stellar collisions place
restrictions on the allowed core configurations of a dense system.  The core
dynamical time is given by (Binney and Tremaine 1987, hereafter BT)

\begin{equation}
t_{\rm dyn,c} \equiv \sqrt{ 3\pi/16G\rho_c } \, ,
\end{equation}

\noindent and the core relaxation time is

\begin{equation}
t_{\rm rel,c} = 0.337 \frac{\sigma^3}{G m_{\star} \rho_c \log\Lambda_c} \, .
\end{equation}

\noindent Here $\rho_c$ is the core density, $\sigma$ is the one-dimensional
velocity dispersion, $m_{\star}$ is the stellar mass,
and $\log \Lambda_c$ is the Coulomb logarithm for the core.  The time scale
for a star in the core to suffer one collision is (BT)

\begin{equation}
t_{col,c} = \left[ 16 \sqrt{\pi} n_c r_{\star}^2 (1+\Theta) \right]^{-1} \, ,
\end{equation}

\noindent where $n_c$ is the stellar number density in the core and
 $r_{\star}$ is the stellar radius.  The Saffronov number $\Theta$ is given by

\begin{equation}
\Theta = \frac{Gm_{\star}}{2 \sigma^{2} r_{\star}} \, .
\end{equation}

Assuming a marginally unresolved 0.37 pc core of solar-mass stars, the central
relaxation time is $6.5\times 10^7$ y, and the collision time scale is
$1.1\times 10^{10}$ y.  These time scales become $3.8\times 10^9$ y and
$6.5\times 10^{11}$ y, respectively, 0.5 parsec from a $3\times 10^6$ \Msun\
black hole in the cusp model of Lauer \etal\ (1992).  For comparison, the
central relaxation time for M13 is $3\times 10^8$ y and the collision time
is $2\times 10^{12}$ years.

There are difficulties with both interpretations of the data.  The short
central relaxation time for the model lacking a central black hole implies
that we have a privileged view to the core collapse of M32.  If the collapse
were reversed by the increased binding energy of binaries, or by the
ejection of stars from the core, then M32 is in postcollapse reexpansion,
and the ``bounce'' event cannot lie in the recent past (GL).  The black hole
interpretation suffers from what GL term the ``luminosity problem'' -- the
complete absence of any activity in the nucleus of M32.  This is
not necessarily fatal to the model either, if M32 is presently between
episodes of flares caused by disrupted stars in the nucleus (Rees 1988).
Presumably, the only unique feature of M32 is its
proximity, so that understanding its core structure will provide insight
into the behavior of similar (but more distant) systems.  We seek therefore
to understand the processes which can lead to M32's remarkable nuclear
properties.

The evolution of stellar systems is usually studied through the use of
Fokker-Planck (hereafter FP) calculations or N-body simulations.  Each
of these techniques possesses strengths and weaknesses.  The principal
advantage which FP calculations enjoy is their ability to handle large-N
systems.  Current N-body techniques can be employed to track up to $10^5$
particles (e.g., Fukushige \etal\ 1991), an order of magnitude lower than
globular clusters and several below small elliptical galaxies.
This is a problem in that the various physical processes which dominate
cluster evolution depend on different powers of N, so there is no simple
way to scale the results from a small-N simulations to a large-N ensemble
(e.g., Giersz and Heggie 1994).  The small-N statistics are often too
noisy for sound analysis.  In addition, the dynamic range of FP
calculations is excellent, whereas in many N-body codes it is limited by
small N or the softening parameter approach to energy conservation.

On the other hand, standard FP calculations make use of an orbit-averaged
diffusion approximation to the system evolution. Thus discrete events, such as
stellar collisions, evolution and mass-loss, must be treated statistically,
whereas they are straightforward to incorporate into N-body codes.  In
addition, the distribution functions used in FP calculations are generally
either functions of $t$ and $E$, or at most $t$, $E$, and $J$, with no
provision made for a possible third integral.  N-body simulations do not
suffer from this problem, since the distribution function need never be
explicitly formulated in terms of conserved integrals.  In addition,
non-axisymmetric mass distributions are at present prohibitively complicated
in an FP calculation, as are evolving off-center nuclear mass concentrations,
whereas in an N-body simulation they require no extra labor.

Nonetheless, there is quite good agreement between FP and N-body simulations
for smaller-N systems under the influence of two-body relaxation.  The
two-body escape rate and core collapse occur in N-body simulations as
predicted by FP calculations (Giersz and Heggie 1994), while provisions made
to include large-angle scattering events in FP calculations have little or
no effect on core collapse (Goodman 1985).  These favorable comparisons make
both techniques valuable for studying dense systems.

An interesting aspect of evolution of a dense rotating stellar system is the
secular evolution of its flattening.  Although globular clusters are quite
different from elliptical galaxies, the fact that the inner regions of some
elliptical galaxies are dense enough to be relaxed makes globular clusters
interesting laboratories wherein to study relaxation processes which
may also operate in the cores of dense ellipticals.  There is observational
evidence to suggest that globular clusters become more spherical with age.
Globulars in the Magellanic Clouds, whose ages range from $10^7$ to $10^{10}$y,
may be somewhat flatter than Galactic globulars, which are thought to be
$10^{10}$ years old (Fall and Frenk 1985).  It has been suggested that
two-body gravitational relaxation in globular clusters transfers angular
momentum to the halo, causing them to become more spherical.  If true we
expect this to hold for any rotating system whose relaxation time
$t_r$ is less than the age of the universe $T_0$.  This fits the general
picture for globular clusters, where $t_r \ll T_0$, and most elliptical
galaxies, where $t_r \gg T_0$.  Agekian (1958) and Shapiro and Marchant (1976)
used Maclaurin spheroids to model a rotating cluster with ellipticity
$\epsilon$, and found that systems with $\epsilon < 0.739$ become more
spherical as they evolve.  Goodman (1983) showed that the Agekian model
predicts that {\it all} flattened clusters become rounder with relaxation
if there is a tidal truncation to the distribution function.  He then analyzed
cluster evolution using a single-mass FP calculation, and showed that cluster
flattening does indeed decrease with time.  Akiyama and Sugimoto (1989) ran
1000 equal-mass particles in an N-body simulation and showed that the angular
momentum contained in cylindrical shells is transported away from the cluster
center, with a characteristic timescale of $~30$ half-mass crossing times.

As energy and angular momentum are transported out of the core  it contracts
and becomes denser, which in turn increases the transport rate, resulting
in a shrinking isothermal core (Lynden-Bell and Wood 1968).  Cohn (1980)
followed this collapse over 20 orders of magnitude in density using a single
mass component FP calculation.  Systems containing a range of masses
evolve much more rapidly than systems with equal-mass stars (Spitzer and
Saslaw 1966; H\'{e}non 1975; Chernoff and Weinberg 1990; Giersz and Heggie
1996); thus we have examined the evolution of systems whose members have a
range of masses as well as those containing stars of equal mass.
 
One possible outcome of unchecked core collapse is an episode of stellar
collisions resulting in the formation of a large black hole (Quinlan and
Shapiro 1990) or a swarm of neutron stars, which in turn would collapse
to form a black hole (Quinlan and Shapiro 1989).  Young (1980) computed
models wherein a pre-existing black hole grows adiabatically at the center
of a spherical cluster.  Lee (1992) used FP calculations to study the
evolution of rotating clusters containing massive black holes in their
centers.  We will delay a discussion of these topics until Paper II, where
we treat stellar collisions, stellar evolution, and stellar mass loss, as
well as the tidal disruption of low-$J$ stars by a central black hole.

It is often argued that the presence of binary stars will affect the evolution
of a stellar system if the density of the system is sufficiently large
(McMillan 1991 and references therein).  The presence of a few ``hard
binaries'' (i.e., binaries whose orbital speed is greater than the local
velocity dispersion) in a globular cluster may be sufficient to halt core
collapse (for a review see Spitzer 1987), but they are unequal to the task
of supporting a larger system like M32.  A pair of 2\Msun\ stars can supply
an amount of energy equal to the total binding energy of a globular cluster.
The binding energy of M32 is perhaps 5 orders of magnitude higher, however,
and so binaries are not likely to affect the evolution of the core of M32.

\section{Initial simulation configuration}

\subsection{Kuzmin-Kutuzov models}

We start each of our simulations with Kuzmin-Kutuzov (hereafter KK) models
(Kuzmin and Kutuzov 1962; Dejonghe and De Zeeuw 1988).  The models are
either spherical, with zero rotation, or rotationally flattened to an axial
ratio of 3.3:1.  The potential-density pair for these models is

\begin{equation}
\Phi(\varpi,z) = -\frac{GM}
{(\varpi^2+z^2+a^2+c^2+2\sqrt{a^2c^2+c^2\varpi^2+a^2z^2})^{1/2}}
\end{equation}

\noindent and

\begin{equation}
\rho(\varpi,z) =
\frac{M c^2}{4\pi} \frac{ (a^2+c^2)\varpi^2 + 2a^2z^2 + 2a^2c^2 + a^4 +
3a^2\sqrt{a^2c^2+c^2\varpi^2 + a^2z^2} }{ (a^2c^2+c^2\varpi^2+a^2z^2)^{3/2}
(\varpi^2 + z^2 + a^2 + c^2 + 2\sqrt{a^2c^2+c^2\varpi^2+a^2z^2})^{3/2} } \, .
\end{equation}

\noindent Here $\varpi$, $z$, and suppressed $\phi$ are cylindrical radius,
height, and azimuth.  The ratio of the scale lengths, $c/a$, corresponds
roughly to the axial ratio (see Dejonghe and De Zeeuw 1988 for a discussion).

The particle velocities are obtained with the KK density/potential pair and
the Jeans equations.  Assuming that the only non-zero first velocity moment
is $v_{\phi}$, and that the model is isotropic (i.e.
$\sigma_{\varpi} = \sigma_{\phi} = \sigma_z$), cross terms like
$\overline{v_{\varpi} v_z}$ disappear from the Jeans equations, and we
obtain

\begin{equation}
\frac{\partial}{\partial \varpi} (\rho \sigma^2) +
\rho \left( -\frac{\overline{v_{\phi}^2}}{\varpi} +
\frac{\partial \Phi}{\partial \varpi} \right) = 0
\end{equation}

\noindent and

\begin{equation}
\frac{\partial}{\partial z} (\rho \sigma^2) =
-\rho \frac{\partial \Phi}{\partial z}
\end{equation}

\noindent To evaluate $\sigma$ at the location of a specific particle we
integrate the last equation over $z$ (Satoh 1980; Binney {\em et al.} 1990),

\begin{equation}
\rho \sigma^2(\varpi ,z) =
\int^{\varpi ,z^{\prime} =\infty}_{\varpi ,z^{\prime} =z}
\rho \frac{\partial \Phi }{\partial z} dz^{\prime}
\end{equation}

\noindent This can be used in to evaluate the rotation speed at a particle
position in the first equation:

\begin{equation}
\rho \frac{\overline{v_{\phi}}^2}{\varpi} = \frac{\partial}{\partial \varpi}
(\rho \sigma^2) + \rho \frac{\partial \Phi}{\partial \varpi}
\end{equation}

\noindent  Figure \ref{vsigcont} shows contours for $\overline{v_{\phi}}$
and $\sigma^2$ for the KK model in which $c/a=0.3$ and $a+c=1$.

\subsection{Distribution sampling}

For a given value of $c/a$ we use the density distribution given above and
perform a rejection-based uniform sampling of a volume of space.  Given a
position $(\varpi,\phi,z)$, we put a particle at that location if

\begin{equation}
\frac{ \rho(\varpi,z) }{ \rho(0,0) } < R
\end{equation}

\noindent where $R$ is a random number on [0,1].  For each particle sampled
at $(\varpi,z)$, we calculate the local dispersion and the particle's rotation
velocity.  The velocity assigned to the particle is simply $\sqrt{3}\sigma$,
oriented randomly in space, plus $\overline{v_{\phi}} \hat{\bf \phi}$.  The
resulting velocities in cylindrical coordinates for 3000 stars are shown in
figure \ref{v3000}.  In the spherical case the velocity distribution for all
three projections is similar to the $(-v_{\varpi},v_z)$ plot.

While this procedure assigns each star a speed on a velocity sphere, the
stars phase mix in a very short time.  It is important to note, however,
that this procedure ensures that the simulated systems do in fact start
out in global virial equilibrium with an isotropic velocity dispersion.
As an example we show various kinematic quantities in cylindrical coordinates
for the rotationally-flattened model II in Table \ref{tc:moments}.  Each of
the averages is calculated globally.  If we compare the binding energy
$V=-0.282551$ to twice the total kinetic energy $2 \times 0.129917$ we get
$(2T+V)/|V|=-8.04$\%.  Although low-N noise contributes to the offset, the
dominant source is the fact that a symmetric distribution about the mean
nearest-neighbor distance produces an asymmetric binding energy distribution
with a negative skew.

\subsection{Initial mass function}

Two mass spectra were employed in these simulations.  Let $P(m) \, dm$
be the probability of a star's mass falling between $m$ and $m+dm$.
The first spectrum used was simply equal-mass stars.  The other spectrum was
a power law.

\begin{equation}
P(m) \, dm = A \, m^{\alpha} \, dm
\end{equation}

\noindent For this case we used $\alpha=-2.35$, the Salpeter initial mass
function (Salpeter 1955).  The spectrum is scaled by choosing a lower and
an upper mass cutoff, $m_{min}$ and $m_{max}$.

The observational literature puts constraints on $m_{min}$ and $m_{max}$.
It is known that a main sequence star must have a mass of at least 0.08\Msun
in order for thermonuclear reactions to proceed in the core (e.g., Grossman,
Hays, and Graboske 1974).  The upper mass limit is somewhat more uncertain.
The most massive star whose mass is directly measured (as a binary member)
is HD47129; its mass is determined to be 30\Msun (Popper 1980).  An
extrapolation of the mass-luminosity relation for an O star in 30 Doradus
yields a mass of about 100\Msun (Humphreys and Davidson 1986).

For the simulations which use a Salpeter IMF we use $m_{min}=0.2$\Msun and
$m_{max}=100$\Msun.  Note that these are limits in the distribution function
to be discretely sampled; an actual collection of 3000 objects will probably
have an upper mass limit much less than 100\Msun, since the probability
density at the high-mass end is very low.  If we have a mass spectrum
characterized by N particles on an interval $[m_{min},m_{max}]$, it is easy
to show that the number of stars N$_i$ on some smaller interval $[m_1,m_2]$ is

\begin{equation}
N_i = N \left(
\frac{m_{2}^{\alpha+1}-m_{1}^{\alpha+1}}{m_{max}^{\alpha+1}-m_{min}^{\alpha+1}}
\right) \, .
\end{equation}

\noindent We thus set $m_{min}$ such that we have at least a star or two on
the subinterval $[25,100]$\Msun\ for N=3000.  Using $m_{min}$=0.2 we expect
N$_i = 3.7 \pm 1.9$ stars on this interval.  A typical sampling of this
distribution is shown in figure \ref{massspec}.  The maximum mass in this set
is 32\Msun.

The mass scale is arbitrary for the simulations in this study, since these
systems evolve under the effects of two-body relaxation only.  Although
results are usually reported in units such that the total system mass is
unity, we occasionally use solar masses to make explicit the observational
constraints on the mass spectrum.  This will also facilitate comparisons with
the simulations of Paper II, wherein the various physical processes treated
require knowledge of the mass scale.

Since stellar evolution and collisions were not allowed, the mass of each
system and the number of stars were constant.  Table \ref{modpars} summarizes
the model parameters for each of the simulations.  In the second column
$a$ and $c$ are the relevant scale lengths in the KK model.  Column 3 refers
to the distribution of stellar masses -- either a collection of equal-mass
stars or the Salpeter IMF -- and columns 4 and 5 give the minimum and maximum
masses, respectively, in the discrete sampling.  Column 6 gives the total
mass of the system in solar masses.

\subsection{Simulations}

To simulate the dynamical evolution of these systems we used the N-body tree
code of Hernquist (1987, 1990), a {\sc Fortran} implementation of the Barnes
and Hut (1986) hierarchical force algorithm.  The simulations were all
performed on a {\sc Sun} workstation.  The output of the program consists
of the mass, position, velocity, energy, and $z$ angular momentum of each
particle, written at specified intervals.  All simulations conserved energy
to better than 0.1\%.

\section{Results}

The evolution in time of quantities presented in this section are plotted
in units of the initial half-mass dynamical time.  As it is instructive to
gauge the progress of various processes with the core and half-mass
relaxation time scales, these quantities are plotted in units of the half-mass
dynamical time in figure \ref{time}.

\subsection{Mass segregation}

As Giersz and Heggie (1996) have noted, the most striking result of
multi-mass model evolution is the rapid core collapse.  Initially
the heavy stars follow the same velocity distribution as the light stars,
and so their temperature is much higher than their surroundings.  They
quickly give up energy to lighter stars and fall into the center of the
system.  It appears that this mechanism operates on roughly the half-mass
relaxation time scale.  In figure \ref{coreIVa} we show the central region of
model IV in $z$ projection at four different times, in center-of-mass
coordinates.  A lump of high-mass stars quickly condenses near the center,
ejecting low-mass stars.

The rapid migration of heavy stars into the core appears to be the result of
the mass segregation instability (Spitzer 1969).  For a system composed of
two mass species (masses $m_1$ and $m_2$, with $m_1<m_2$; total mass in each
species $M_1$ and $M_2$), Spitzer (1969, 1987) defines the stability parameter
$\chi$ as

\begin{equation}
\chi = \frac{M_2}{M_1} \left(\frac{m_2}{m_1}\right)^{3/2}
\end{equation}

\noindent for $m_2\gg m_1$ and $M_2\ll M_1$.  If
$\chi < \chi_{\rm crit} = 0.16$ then the system is stable against runaway
mass segregation.  If not, the self-gravity of the heavy stars requires
a velocity dispersion that prevents equipartition with the lighter stars.
The heavy stars give up energy to the lighter stars and sink to the center,
while the light stars carry this energy into the halo.  Since our simulations
employed stars of many different masses we chose a separation mass
$m_{\rm sep}$ to categorize stars as either $m_1$ or $m_2$.  That is, if
$m_\star < m_{\rm sep}$ then we treat it as type $m_1$, and if
$m_\star > m_{\rm sep}$ then we treat it as type $m_2$.  For every value of
$m_{\rm sep}$ chosen, $\chi \gg \chi_{\rm crit}$.  Thus our systems were
unstable to runaway mass segregation.

Following the work of King (1966) and the argument of Giersz and Heggie
(1996) we define the ``core radius'' of the system as

\begin{equation}
  r_c^2 = \frac{9 \sigma_c^2}{4\pi G \rho_c} \, ,
\end{equation}

\noindent where $\sigma_c$ and $\rho_c$ are the central velocity dispersion
and density.  Giersz and Heggie calculated the central quantities over
the innermost 1\% of the mass of the system, without worrying about the noise
this incurs because of the simulation-averaging technique they employed.
Since we ran only one simulation for each set of parameters we used the
innermost 5\% of the system mass in our core calculations.  Although this
produces slightly larger core radii, it is the {\it behavior} of the core
radius defined at a particular level, and those quantities dependent upon it,
which are of interest.

Defined in this way, the core radius plummets in about a half-mass relaxation
time, or 10 half-mass crossing times, for multi-mass models III and IV (see
figure \ref{corePmm}).  Once the core collapse phase has ended, the
resulting core size depends on the softening length.  Figure \ref{core_comp}
shows the evolution of the core radius for softening lengths of 0.1 and 0.025.

Our simulations show that core collapse slows as low-mass stars are
evacuated from the core, consistent with the results of Giersz and Heggie
(1996).  Their depletion removes the dominant mechanism whereby energy is
transported from the core to the halo (Spitzer 1969, 1987).  Figure
\ref{coreIVb} shows the $z$ projection of model IV for $r<0.20$ and $r>0.20$,
at $t/t_{\rm dyn,h}=102$.  Here $r$ is measured from the lump center, defined
as the mass center of the 10 stars with the lowest potential energy.
Global mass segregation is shown in figure \ref{massdist}, which plots the
characteristic scale length $(\overline{1/r})^{-1}$ for different mass groups
as a function of time.

Once the mass concentration forms it moves slowly around the central part
of the system as defined by the stars not bound to it.  The motion of the
``density center'' of a stellar system has been noted by several groups
(although they have defined it in different ways).  Miller and Smith (1992)
described the behavior of the density center of their simulations as an
``oscillation'' and a ``growing wave.''  Sweatman (1993), in contrast,
argued that the motions are mainly a $1/\sqrt{N}$ noise effect.  Spurzem
and Aarseth (1996) Fourier analyzed the motion of the lump in their
simulations and showed that it orbital timescale was about 14 half-mass
crossing times (quite consistent with the timescale we observed).  All of
these studies involved systems of equal-mass stars.  While we observe a
similar phenomenon in our equal-mass simulations, it is much more striking
in the mass-spectrum simulations.  In these experiments the membership of 
the lump is nearly constant.

A density enhancement also forms at the center of our equal-mass
systems, but it has a somewhat more evanescent character, losing and
accreting members.  We believe that this is the same phenomenon, but
it is certainly much less striking.  It therefore seems likely that
mass segregation plays an important role in the strength of the
phenomenon.  We believe that, whether long-lived or not, the behavior 
of the lump is Brownian motion due to scattering of individual stars.

Figure \ref{lumpcm} shows
the position of the model IV lump at 16 equally-spaced time intervals spanning
the simulation.  Using our $r\leq 0.2$ lump membership definition, we
replaced all the lump members by a single particle with the former lump's
mass and velocity.  We then advanced the simulation by another 50 half-mass
crossing times.  The position of the lump replacement is shown in figure
\ref{lumprep}.  The initial and final positions, (0.11,0.10) and (0.19,0.61),
respectively, have been circled.  The initial velocity vector of the particle
is in the positive $x$ and negative $y$ direction.

The kinetic energy of the lump is

\begin{equation}
T_{\rm lump} = \frac{1}{2} M_{\rm lump} |{\bf V}_{\rm lump}|^2 \, .
\end{equation}

\noindent We compare this quantity to the average kinetic energy per
particle of the lump environment, defined as a cumulative average
of the kinetic energies of particles not found within in the lump sphere:

\begin{equation}
T_{\rm env}(r) = \frac{ {\displaystyle \sum_{i}^{N_{env}} } \,
m_i v_i^2}{2 N_{\rm env}} \, .
\end{equation}

A typical environment temperature profile is shown in figure \ref{lumpT}.
$T_{\rm env}$ is plotted as a function of distance from the system
center-of-mass, for model IV at $t/t_{\rm dyn,h}=102$.  The lump temperature
at this time is about 0.0011; model IV in general shows a lump temperature in
the range $0.006 < T < 0.0012$ and likewise for the environment temperature at
low $r$.  Although the low number of particles makes for noisy statistics, the
evidence here suggests that the lump is in Brownian motion about the central
region of the system, in thermal equilibrium with the immediate environment.

% HERE

\subsection{Flattening}

An interesting aspect of the evolution of rotating systems is the change
in the flatness of the system, as a function of both radius and time.
One way to characterize this is to fit ellipses to projected light or density
contours.  The ellipticity is then defined as

\begin{equation}
\epsilon = 1 - b/a \, ,
\end{equation}

\noindent where $a$ and $b$ are the major and minor semiaxes of the ellipse.
Another way is to use the ``dynamical ellipticity'' (Goodman 1983), which
characterizes the flattening as being due to rotation.  It assumes that the
three-dimensional density contours are concentric spheroids, and that the
distribution function is a function of $E$ and $J$ only.  While this definition
is useful for FP calculations in which $f=f(E,J)$, it is not appropriate for
N-body simulations, where no assumptions about the distribution function are
made.

We seek to characterize the flatness of each system in a way that is
numerically simple and that makes only minimal assumptions about the mass
distribution.  Defining $\theta$ as {\it latitude}, we adopt

\begin{equation}
\epsilon = 1 - \langle | \tan\theta | \rangle \, ,
\end{equation}

\noindent where $\langle | \tan\theta | \rangle $ is the
density-weighted average of $|\tan\theta|$.  The flatness of a thick
{\it spherical} shell bounded by $r_1$ and $r_2$ is thus defined as

\begin{equation}
\epsilon = 1 -
\frac{ {\displaystyle \int^{r_2}_{r_1}} {\displaystyle \int^{\pi/2}_{-\pi/2}}
\, {\displaystyle \int^{2\pi}_{0}}
\, |\tan\theta| \, \rho(r,\theta,\phi) \, r^2 \cos\theta \, dr \, d\theta
\, d\phi}
{ {\displaystyle \int^{r_2}_{r_1}} {\displaystyle \int^{\pi/2}_{-\pi/2}}
\, {\displaystyle \int^{2\pi}_{0}}
\, \rho(r,\theta,\phi) \, r^2 \cos\theta \, dr \, d\theta \, d\phi} \, .
\end{equation}

\noindent We now assume $z$ axisymmetry and convert to a sum over discrete
particles:

\begin{equation}
\epsilon = 1 - \frac{{\displaystyle \sum_i^{N_{shell}}} \, m_i \,
|\tan\theta_i|} {{\displaystyle \sum_i^{N_{shell}}} \, m_i} \, .
\end{equation}

\noindent Here the summations are performed over all particles found within
the spherical shell of interest.  For oblate systems $\epsilon$ lies on
$(0,1]$, while for prolate systems it spans $[-\infty,0)$; for a sphere
$\epsilon=0$.  Tests of this statistic can be found in the appendix.

Consistent with the FP results of Goodman (1983), we find that our flattened
models become secularly rounder.  We plot ellipticity versus time for
Lagrangian quintiles of model II in figure \ref{ellipII} and of model IV in
figure \ref{ellipIV}.  It is clear that the system becomes significantly
less flat in the inner 50\% of the system.  As expected, the decrease in
flattening is more pronounced in the inner quintiles, where the local
relaxation time scale is comparatively short.

It appears that the reduction is flattening is reduced somewhat
for the multi-mass model.  While the flattening continues to decrease up
to $t=280 t_{dyn,h}$ in the equal-mass simulation, the reduction tails off
around $t=150-200 t_{dyn,h}$.  This is probably due to the fact that light
stars are ejected to the the halo rapidly during the early phases of the
evolution as the heavy stars become centrally condensed; they gain energy
more rapidly than they can lose angular momentum.

A crude way to see the trend more clearly is to make a least-squares fit to
the Lagrangian quintile ellipticities for all the models.  Figure
\ref{ellipt} shows this for the four models.  In the flattened models, the
outer 80\% by mass of each system becomes rounder with time.  The linear
fits for the innermost quintile are not statistically significant for the
Salpeter IMF models since the number of particles in the innermost quintile
is low, resulting in excessive noise.

It appears that the the inner region of the flattened equal-mass simulation
becomes more spherical than the multi-mass simulation.  We suspect this is
due to a difference in the angular momentum transport in these regions
mentioned above.  In figure \ref{J} we show the angular momentum per unit
mass for the inner three quintiles of the rotating models.  Model II, the
equal-mass system, clearly transports more angular momentum out of this
region than does model IV.

\section{Summary}

In this study we have examined the evolution of dense stellar systems using
N-body simulations.  The results obtained concur with Giersz and Heggie
(1996) -- we find that systems whose stars have many different masses undergo
rapid core collapse.  The high-mass stars quickly condense into a dyamically
distinct lump at the center.  This core collapse terminates when the vast
majority of low-mass stars, the carriers in the energy transport which
produces core collapse, has been ejected from nucleus.  The size of the lump
is limited by the value of the softening parameter.  We have determined that
the lump, once formed, wanders about the the nucleus in Brownian motion, in
thermal equilibrium with the stars in the nucleus of the system.

We have devised a statistic to characterize the flatness of a system.
We found that rotationally flattened systems become less flat in time,
concurring with the FP results of Goodman (1983).  This decrease in
flattening is more pronounced in the inner regions of each system, where
the relaxation time is comparatively short.  We have found that a
multi-mass system retains its flatness somewhat better than an equal-mass
system  over the inner half of their mass.  This is probably related to
the fact that the equal mass system transports more angular momentum
out of its core than does the multi-mass system.

\acknowledgments
This research was supported by NASA Theory Grant NAG 5-2758.

\appendix
\section{Tests of the flattening statistic}

While the flattening statistic $\epsilon$ defined in the text is
computationally simple, it is not obvious that it is a good indicator of the
``true'' flatness of the system.  Noise might be a problem in systems with
only a few thousand particles, and the mathematical definition of $\epsilon$
is formally singular if a particle crosses the $z$ axis.  In addition, while a
radial modulation in density is necessary for any flattening to register, it
will also introduce a systematic bias in the calculation of $\epsilon$.  For
example, we might expect our prescription to artificially inflate the value of
the flattening if $-d\log\rho/d\log r$ is large, since we would be biasing any
mass sample with material at the equator.  We investigated the validity of
this prescription for computing $\epsilon$ in two ways.  The first was an
analytic calculation of $\epsilon$ for an infinitesimally thin spherical shell
in an ellipsoidal scale-free system.  We assumed a density distribution of the
form

\begin{equation}
\frac{\rho(\varpi,z)}{\rho_0} =
\left( \frac{r_0}{\sqrt{\varpi^2+\alpha^2 z^2}} \right)^{\beta} \, .
\label{eq:rho}
\end{equation}

\noindent Here the parameter $\alpha$ is related to the intrinsic
ellipticity $\epsilon_{\rm int}$ by

\begin{equation}
\epsilon_{\rm int} = 1 - \frac{1}{\alpha} \, .
\end{equation}

\noindent The amount of material $\delta M$ contained in a thin shell of
radius $r$ and width $\delta r$ is

\begin{equation}
\delta M = \frac{2\pi \rho_0 r_0^{\beta} \delta r}{r^{\beta-2}}
\int_{-\pi/2}^{\pi/2} \,
\frac{\cos\theta d\theta}{(\cos^2\theta+\alpha^2\sin^2\theta)^{\beta/2}} \, .
\end{equation}

\noindent The calculated flatness is

\begin{equation}
\epsilon = 1 - \frac{ {\displaystyle \int_{-\pi/2}^{\pi/2}} \, |\tan\theta| \,
\frac{\cos\theta d\theta}{(\cos^2\theta+\alpha^2\sin^2\theta)^{\beta/2}} }
{ {\displaystyle \int_{-\pi/2}^{\pi/2}} \,
\frac{\cos\theta d\theta}{(\cos^2\theta+\alpha^2\sin^2\theta)^{\beta/2}} } \, .
\end{equation}

\noindent For even integral $\beta>2$ we have

\begin{equation}
\epsilon \, = \, 1 \, - \, \frac{
{\displaystyle \sum_{j=1}^{\beta/2-1} }
\frac{j!(j-1)!}{\alpha^2(2\alpha)^{\beta-2-2j}(2j)!} \, + \,
\frac{1}{\alpha(2\alpha)^{\beta-2}\sqrt{\alpha^2-1}} \cdot
\log\left(
\frac{\alpha+\sqrt{\alpha^2-1}}{\alpha-\sqrt{\alpha^2-1}}
\right)
}{
{\displaystyle \sum_{j=1}^{\beta/2-1}}
\frac{j!(j-1)!}{2^{\beta-2-2j}(2j)!\,\alpha^{2j}} \, + \,
\frac{1}{2^{\beta-1}\sqrt{\alpha^2-1}} \cdot \tan^{-1}(\sqrt{\alpha^2-1})
} \, .
\end{equation}

\noindent Figure \ref{flat_anal} shows the calculated flatness versus the
intrinsic ellipticity for three values of $\beta$ over the entire range of
oblate objects.  As expected, the bias introduced by this method is directed
toward a greater flattening, although the amount is not very significant.  In
the outer regions of a KK model $\beta=4$, so we do not expect significant
errors.  Even for the halo of a flattened Plummer law, where $\beta=5$, the
error would still not be intolerably large.

Thus the second way we examined this statistic was through the use of Monte
Carlo simulations.  We sampled the power law mass distribution given by
equation \ref{eq:rho} for 3000 particles outside of some minimum radius
$r_{\rm min}$.  We then divided the system into five equal-mass spherical
shells and calculated the flattening of each.  For each value of $\beta$ and
$\epsilon_{\rm int}$ (setting $r_{\rm min}=1$) we performed 100 experiments,
finding the mean value of $\epsilon$ and the standard deviation for each shell.
The results are plotted in figure \ref{flat_mc}.  Consistent with our analytic
calculations, the calculated flatness overestimates the intrinsic flattening,
with the discrepancy increasing with increasing $\beta$.  Kuzmin-Kutuzov
models decay as $\beta=4$ or shallower, however, so the error is not large.

\begin{center}
REFERENCES
\end{center}

\def\ARAA{{\it Ann. Rev. Astron. Astrophys.}}
\def\ApJ{{\it Ap. J.}}
\def\ApJL{{\it Ap. J. Lett.}}
\def\ApJSS{{\it Ap. J. Supp. Ser.}}
\def\AandA{{\it Astron. Astrophys.}}
\def\AJ{{\it Astron. J.}}
\def\JCP{{\it J. Comp. Phys.}}
\def\MNRAS{{\it M. N. R. A. S.}}
\def\N{{\it Nature}}
\def\PASJ{{\it Publ. Astron. Soc. Jap.}}
\def\RPP{{\it Rep. Prog. Phys.}}

\begin{verse}

Agekian, T. A. 1958, {\it Soviet Astronomy A. J.}, {\bf 2}, 22.

Akiyama, K. and Sugimoto, D. 1989, \PASJ, {\bf 41}, 991.

Barnes, J., and Hut, P. 1986, \N, {\bf 324}, 446.

Binney, J. J., Davies, R. L., and Illingworth, G. D. 1990, \ApJ {\bf 361}, 78.

Binney, J. J., and Tremaine, S. 1987, {\it Galactic Dynamics} (Princeton:
Princeton University Press).

Chernoff, D. F., and Weinberg, M. D. 1990, \ApJ, {\bf 351}, 121.

Cohn, H. N. 1980, \ApJ {\bf 242}, 765.

Dejonghe, H., and De Zeeuw, T. 1988, \ApJ, {\bf 333}, 90.

Djorgovski, S. 1993, in {\it ASP Conference Series 50, Structure
and Dynamics of Globular Clusters}, ed. S. G. Djorgovski and G. Meylan
(San Francisco: Astronomical Society of the Pacific), 373.

Djorgovski, S., and King, I. R. 1986, \ApJL, {\bf 305}, L61.

Dressler, A., and Richstone, D. O. 1988, \ApJ, {\bf 324}, 701.

Fall, S. M., and Frenk, C. S. 1985, in {\it IAU Symp. No. 113, Dynamics of
Star Clusters}, ed. J. J. Goodman and P. Hut (D. Reidel Publishing Company,
Dordrecht), p. 285.

Frenk, C. S., and Fall, S. M. 1982, \MNRAS, {\bf 199}, 565.

Fukushige, T., Ito, T., Makino, J., Ebisuzaki, T., Sugimoto, D., and Umemura,
M. 1991, \PASJ, {\bf 43}, 841.

Giersz, M., and Heggie, C. D. 1994, \MNRAS, {\bf 268}, 257.

Giersz, M., and Heggie, C. D. 1996, \MNRAS, {\bf 279}, 1037.

Giersz, M., and Heggie, C. D. in {\it IAU Symp. No. 174, Dynamical Evolution
of Star Clusters}, ed. P. Hut and J. Makino (Kluwer, Dordrecht), in press.

Goodman, J. 1983, Ph.D. Thesis, Princeton University.

Goodman, J. 1985, in {\it IAU Symp. No. 113, Dynamics of Star Clusters}, ed.
J. J. Goodman and P. Hut (D. Reidel Publishing Company, Dordrecht), p. 285.

Goodman, J. and Lee, H. M. 1989 \ApJ, {\bf 337}, 84.

Grossman, A. S., Hays, D., and Graboske, H. C. 1974, \AandA, {\bf 30}, 95.

H\'{e}non, M. 1975, in {\it IAU Symp. No. 69, Dynamics of Stellar Systems},
ed. A. Hayli (D. Reidel Publishing Company, Dordrecht), p. 133.

Hernquist, L. 1987, \ApJSS, {\bf 64}, 715.
 
Hernquist, L. 1990, \JCP, {\bf 87}, 137.

Humphreys, R. M., and Davidson, K. 1986, {\it New Scientist}, {\bf 112}, 38.

King, I. R. 1966, \AJ, {\bf 71}, 64.

Kuzmin, G. G., and Kutuzov, S. A. 1962, {\it Bull. Abastumani Ap. Obs.},
{\bf 27}, 82.

Lauer, T., Faber, S. M., Lynds, C. R., Baum, W. A., Ewald, S. P., Groth,
E. J., Hester, J. J., Holtzman, J. A., Kristian, J., and Light, R. M. 1992,
\AJ, {\bf 103}, 703.

Lee, M. H. 1992, Ph.D. Thesis, Princeton University.

Lynden-Bell, D., and Wood, R. 1968, \MNRAS, {\bf 138}, 495.

McMillan, S. L. W. 1991, in {\it ASP Conference Series 13, The Formation and
Evolution of Star Clusters}, ed. K. Janes (San Francisco: Astronomical Society
of the Pacific), p. 324.

Miller, R. H., and Smith, B. F. 1992, \ApJ, {\bf 393}, 508.

Popper, D. M. 1980, \ARAA, {\bf 18}, 115.

Qian, E. E., De Zeeuw, T., van der Marel, R. P., and Hunter, C. 1995, \MNRAS,
{\bf 274}, 602.

Quinlan, G. D., and Shapiro, S. L. 1990, \ApJ, {\bf 356}, 483.

Quinlan, G. D., and Shapiro, S. L. 1989, \ApJ, {\bf 343}, 725.

Rees, M. J. 1988, \N, {\bf 333}, 523.

Richstone, D. O., Bower, G., and Dressler, A. 1990, \ApJ, {\bf 353}, 118.

Salpeter, E. E. 1955, \ApJ, {\bf 121}, 161.

Satoh, C. 1980, \PASJ, {\bf 32}, 41.

Shapiro, S. L., and Marchant, A. B. 1976, \ApJ, {\bf 210},757.

Spitzer, L. Jr. 1987, {\it The Dynamical Evolution of Globular Clusters}
(Princeton: Princeton University Press).

Spitzer, L. Jr. 1969, \ApJL, {\bf 158}, L139.

Spitzer, L. Jr., and Saslaw, W. C., 1966, \ApJ, {\bf 143}, 400.

Spurzem, R., and Aarseth, S. J. 1996, \MNRAS, {\bf 282}, 19.

Sweatman, W. L. 1993, \MNRAS, {\bf 261}, 497.

van der Marel, R. P., de Zeeuw, P. T., Rix, H. W., and Quinlan, G. D. 1997,
\N, {\bf 385}, 610.

van der Marel, R. P., Evans, N. W., Rix, H. W., White, S. D. M., and De Zeeuw,
T. 1994a, \MNRAS, {\bf 271}, 99.

van der Marel, R. P., Rix, H. W., Carter, D., Franx, M., White, S. D. M.,
and De Zeeuw, T. 1994b, \MNRAS, {\bf 268}, 521.

Young, P. 1980, \ApJ {\bf 242}, 1232.

\end{verse}

%  TABLES

\clearpage
\begin{deluxetable}{cccc}
\tablecaption{
Cylindrical velocity moments for the initial configuration of model II. This
system is rotationally flattened to $c/a=0.3$. \label{tc:moments}}
\tablehead{
\colhead{coordinate}         &
\colhead{$\overline{v_i}$}   &
\colhead{$\overline{v_i^2}$} &
\colhead{$\sigma_i$}
}
\startdata
   $\varpi$   &    0.00317032    &     0.04847563     &  0.22014898  \nl
   $\phi$     &    0.32752717    &     0.16131656     &  0.23247047  \nl
   $z$        &    0.00576263    &     0.05004266     &  0.22362792  \nl
\enddata
\end{deluxetable}

\clearpage

\begin{deluxetable}{lclccc}
\tablecaption{
Initial model parameters for the four N=3000 simulations in this study.
\label{modpars}}
\tablehead{
\colhead{model}                  &
\colhead{$1-c/a$}                &
\colhead{$N(m)$}                 &
\colhead{$m_{\rm smallest}$/\Msun}    &
\colhead{$m_{\rm largest}$/\Msun}    &
\colhead{M$_{\rm system}$/\Msun}
}
\startdata
I    & 0.0 & $\delta(m-1M_{\odot})$ & 1.000 & 1.000 & 3000.0 \nl
II   & 0.7 & $\delta(m-1M_{\odot})$ & 1.000 & 1.000 & 3000.0 \nl
III  & 0.0 & $m^{-2.35}$            & 0.200 & 32.01 & 1890.7 \nl
IV   & 0.7 & $m^{-2.35}$            & 0.200 & 32.01 & 1890.7 \nl
\enddata
\end{deluxetable}

% FIGURES and CAPTIONS

\clearpage
\plotone{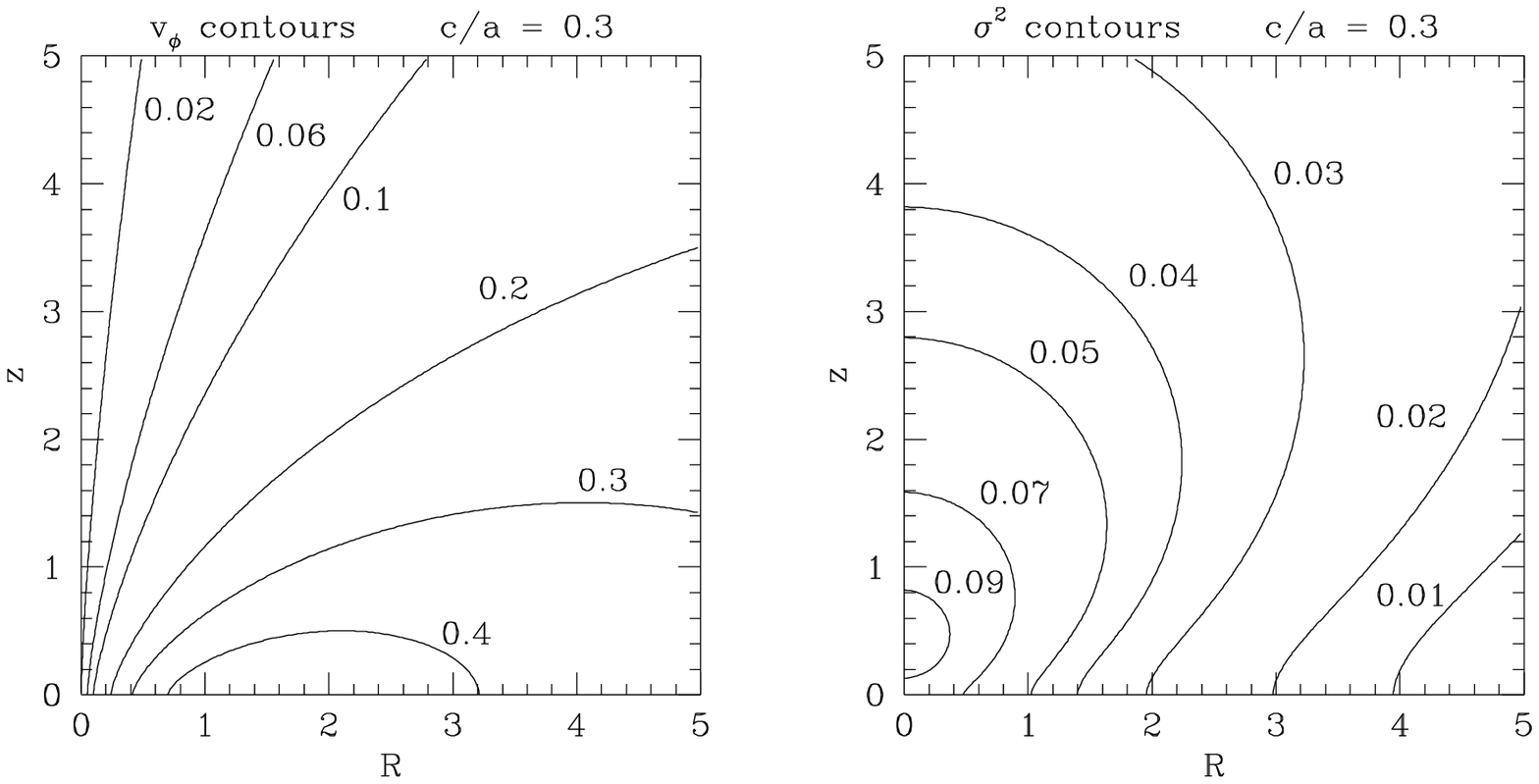}

\figcaption{Contours in $v_{\phi}$ and $\sigma^2$ for a flattened
Kuzmin-Kutuzov model. R plotted here is $\varpi$ in the text.
\label{vsigcont}}

\clearpage
\plotone{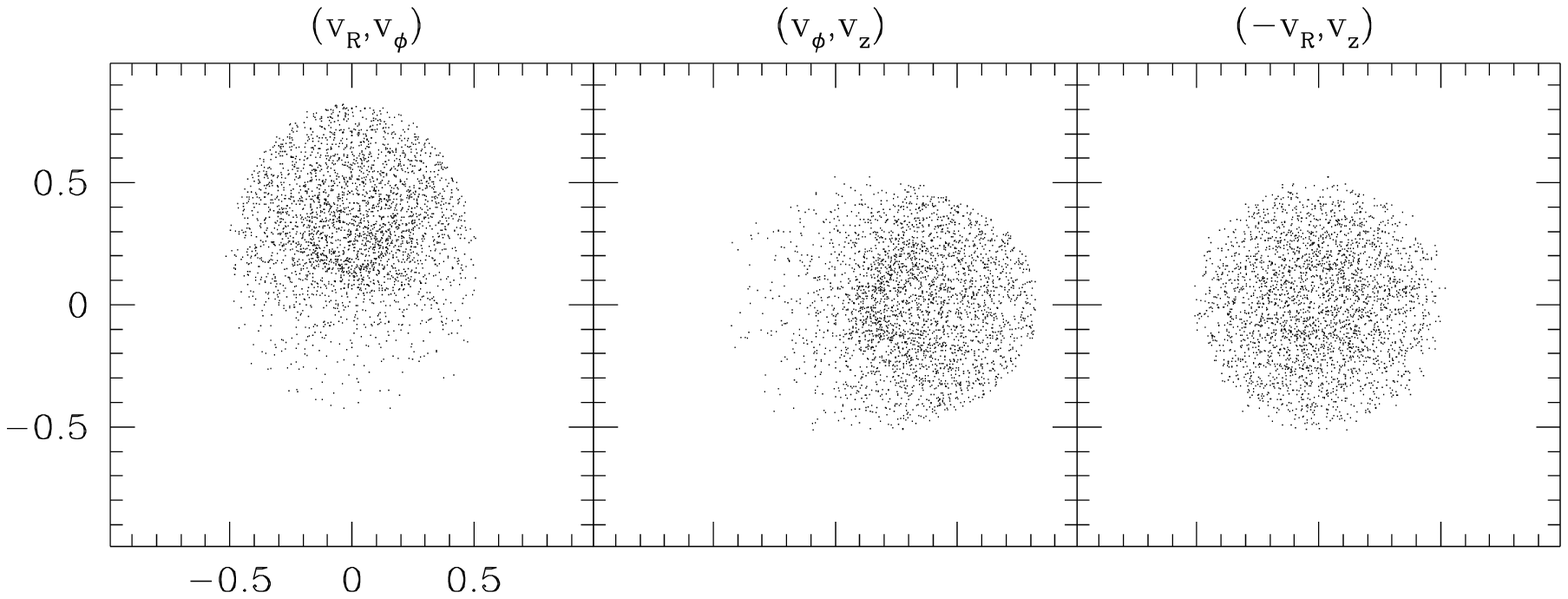}

\figcaption{Stellar velocities for 3000 stars sampled from a $c/a=0.3$
Kuzmin-Kutuzov model. \label{v3000}}

\clearpage
\plotone{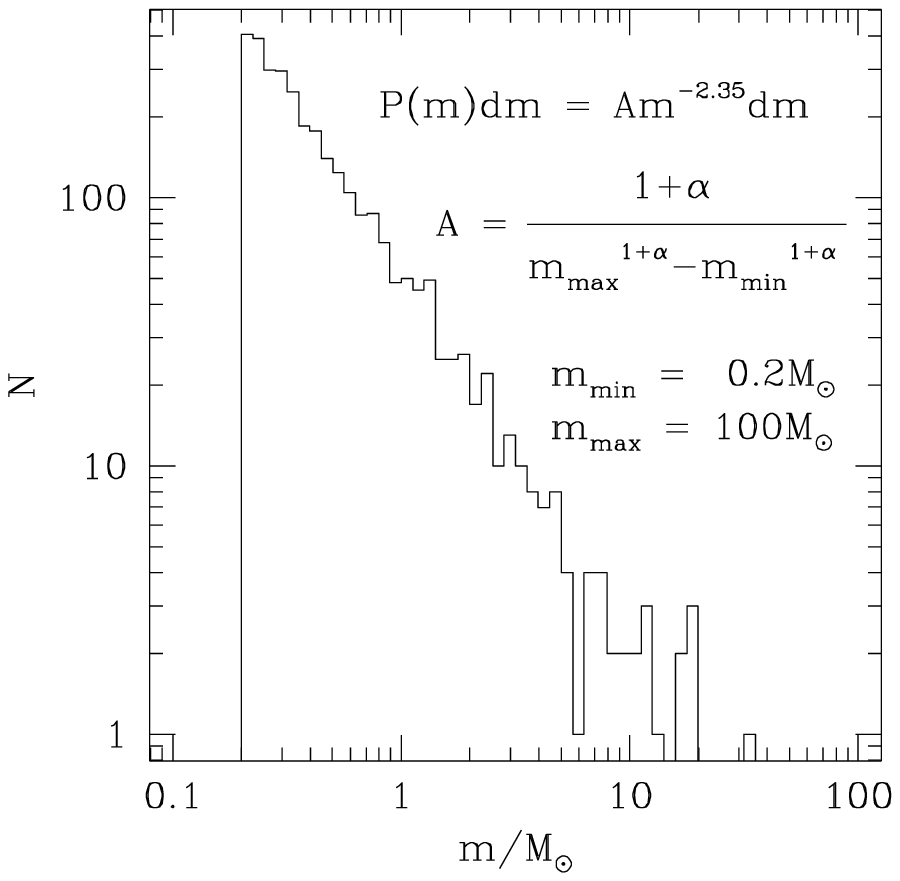}

\figcaption{A Salpeter spectrum sampled for 3000 bodies on [0.2,100]\Msun.
\label{massspec}}

\clearpage
\plotone{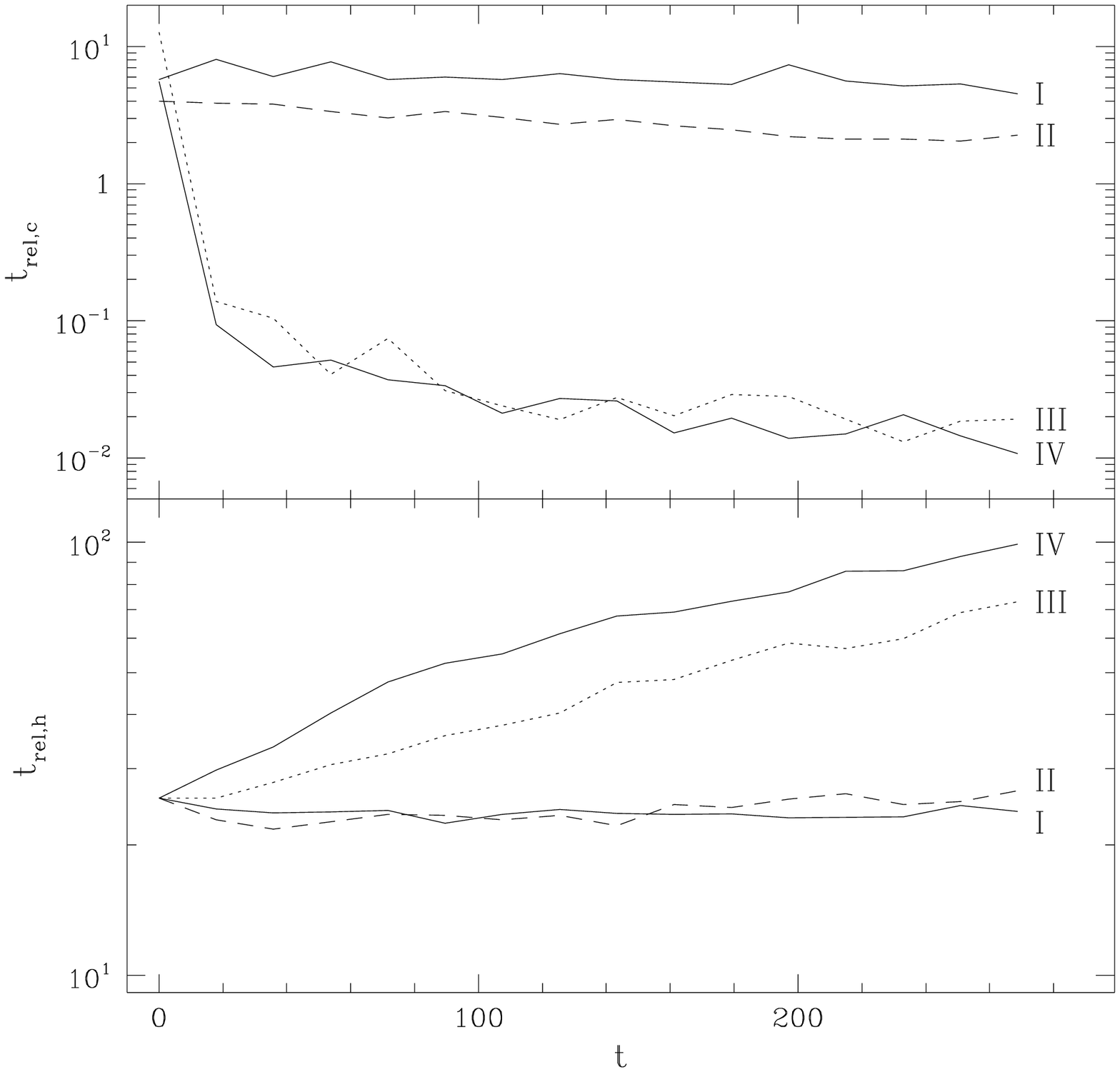}

\figcaption{Core and half-mass relaxation times for the 4 models.  All
quantities are plotted in units of the initial half-mass dynamical time scale.
\label{time}}

\clearpage
\plotone{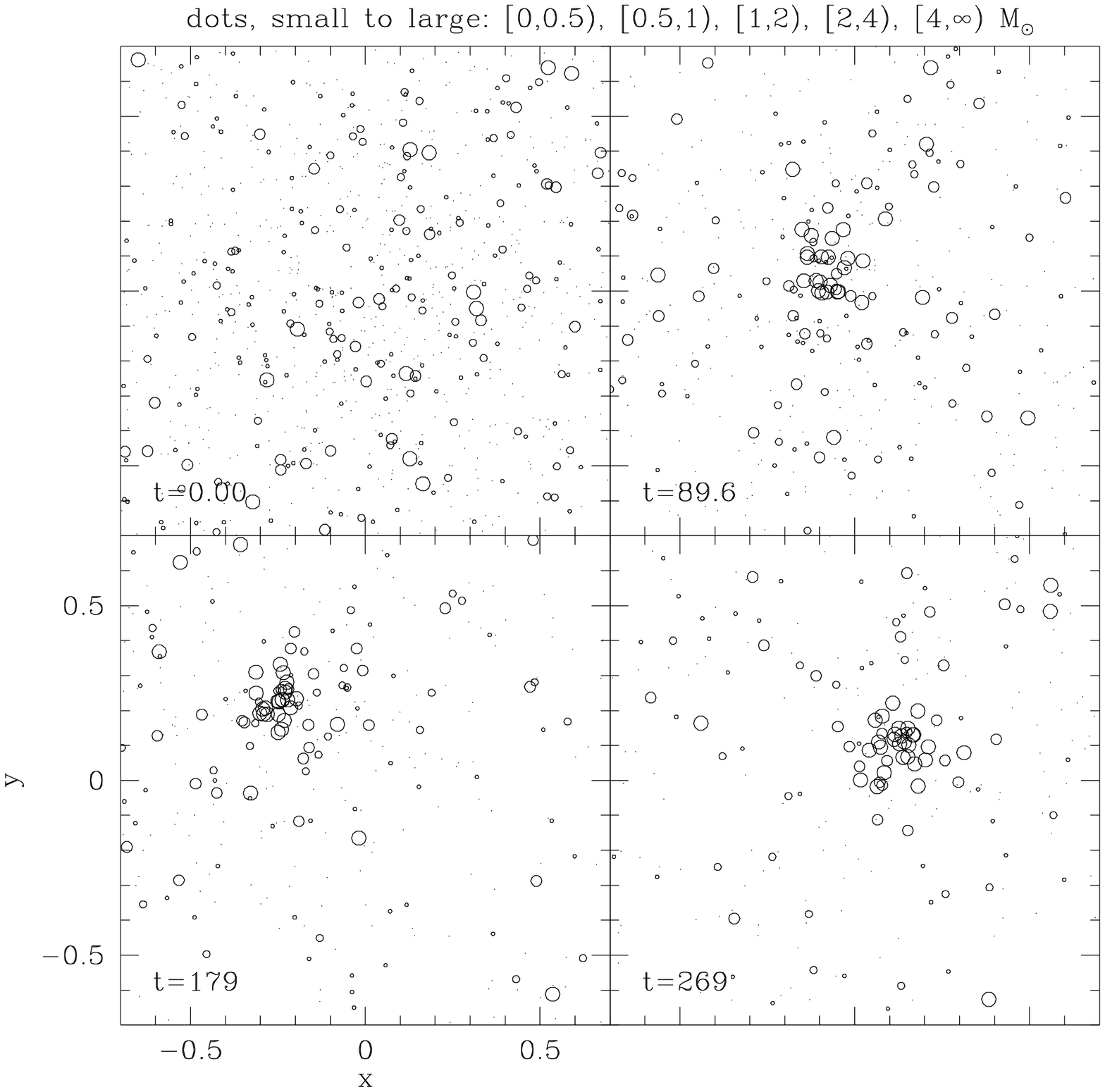}

\figcaption{Z-projection of the core of model IV ($t$ in units of the
half-mass dynamical time).\label{coreIVa}}

\clearpage
\plotone{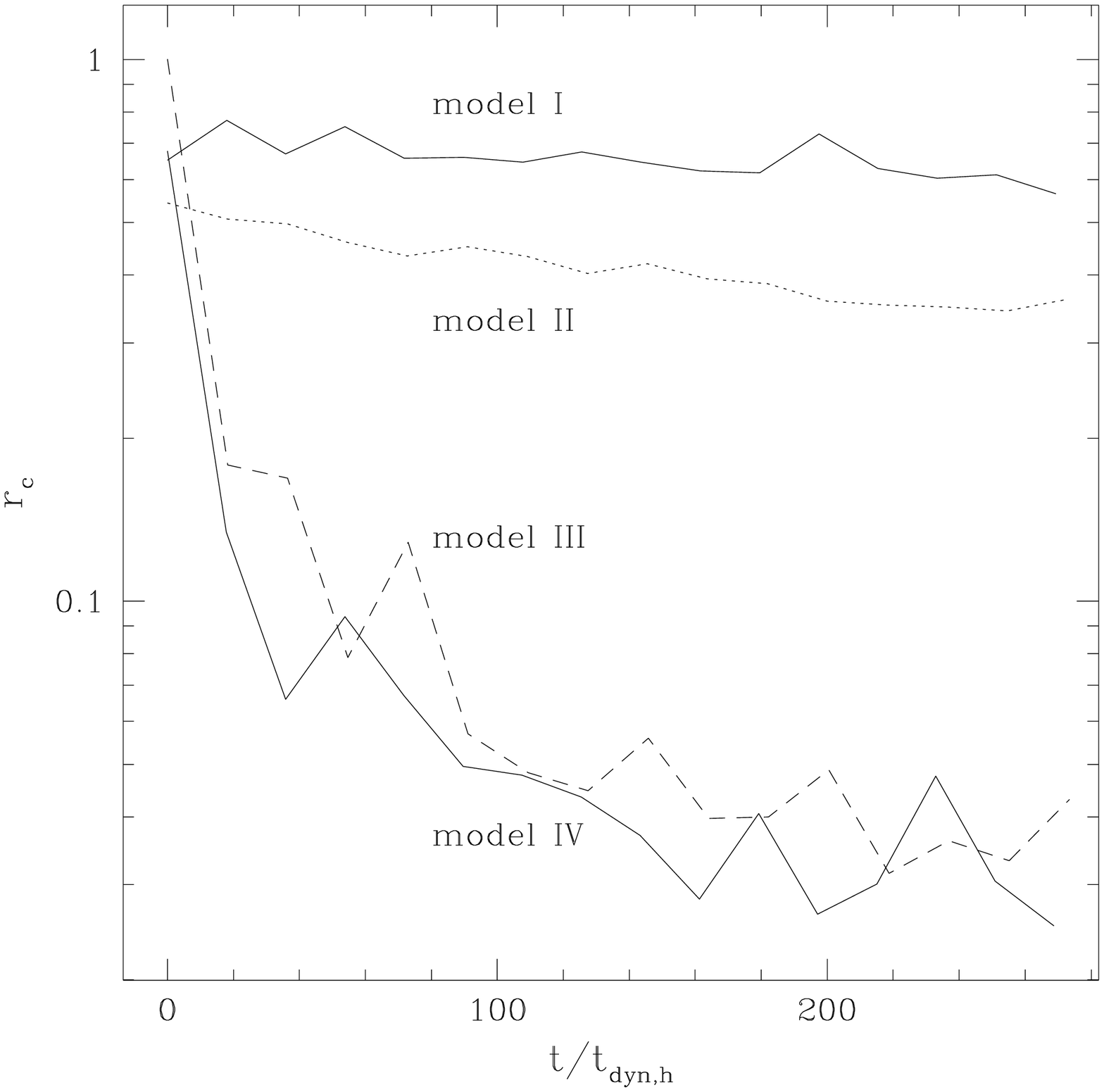}

\figcaption{Core radius versus time for the four simulations.
\label{corePmm}}

\clearpage
\plotone{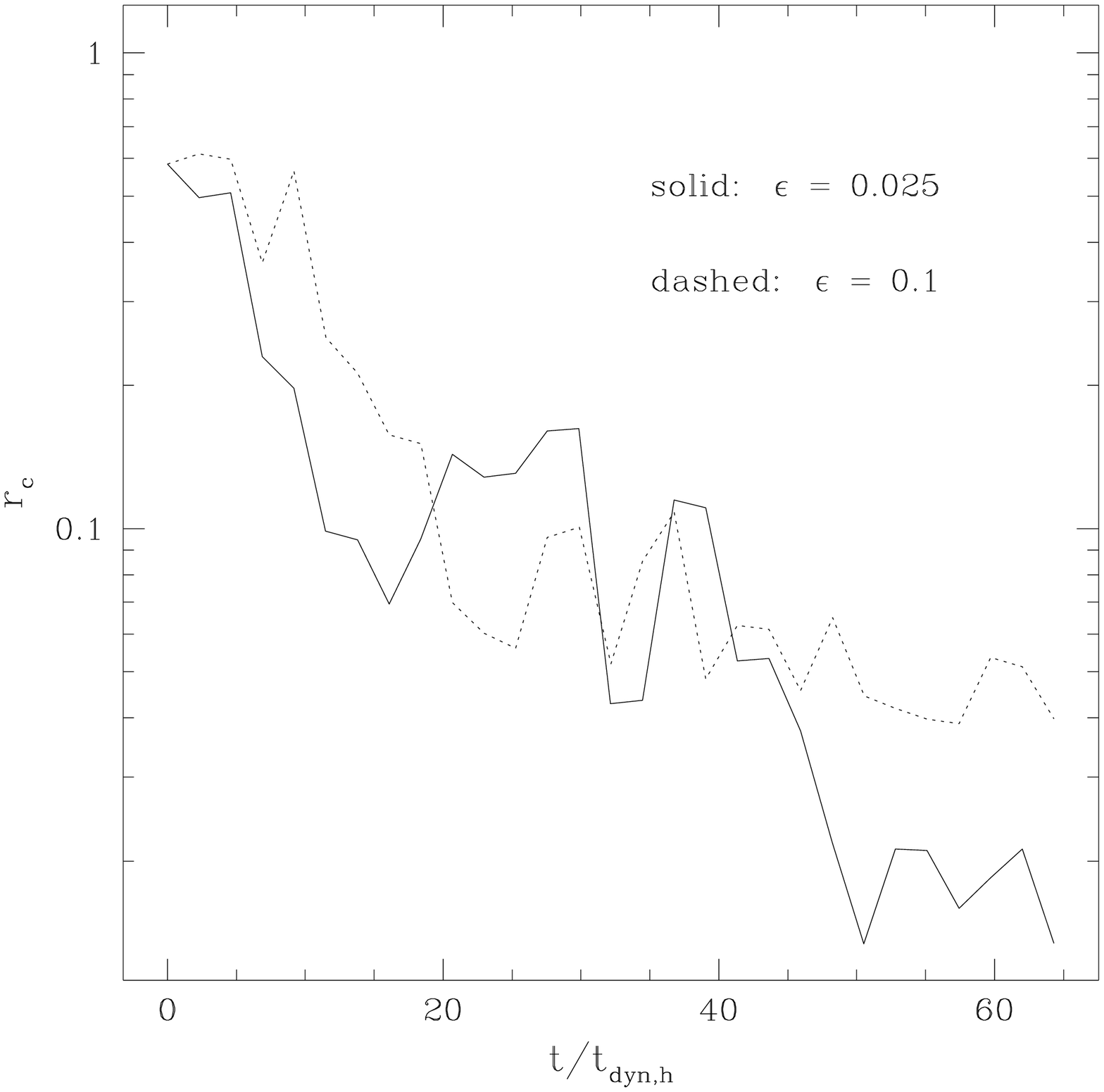}

\figcaption{Core collapse for two different values of the softening length.
\label{core_comp}}

\clearpage
\plotone{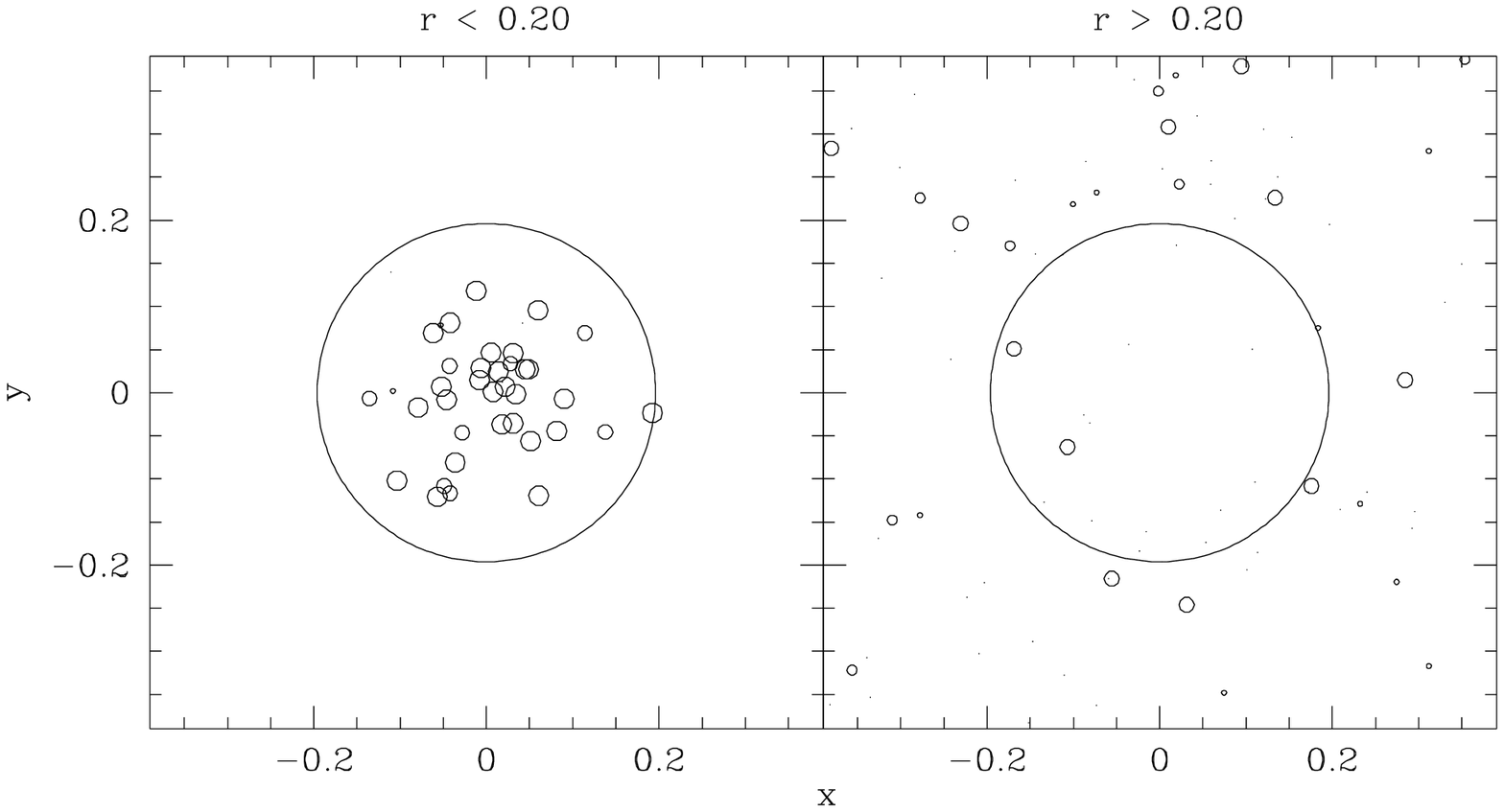}

\figcaption{z-projection of the core region of model IV for $r<0.2$ and
$r>0.2$ at $t/t_{\rm dyn,h}=102$. \label{coreIVb}}

\clearpage
\plotone{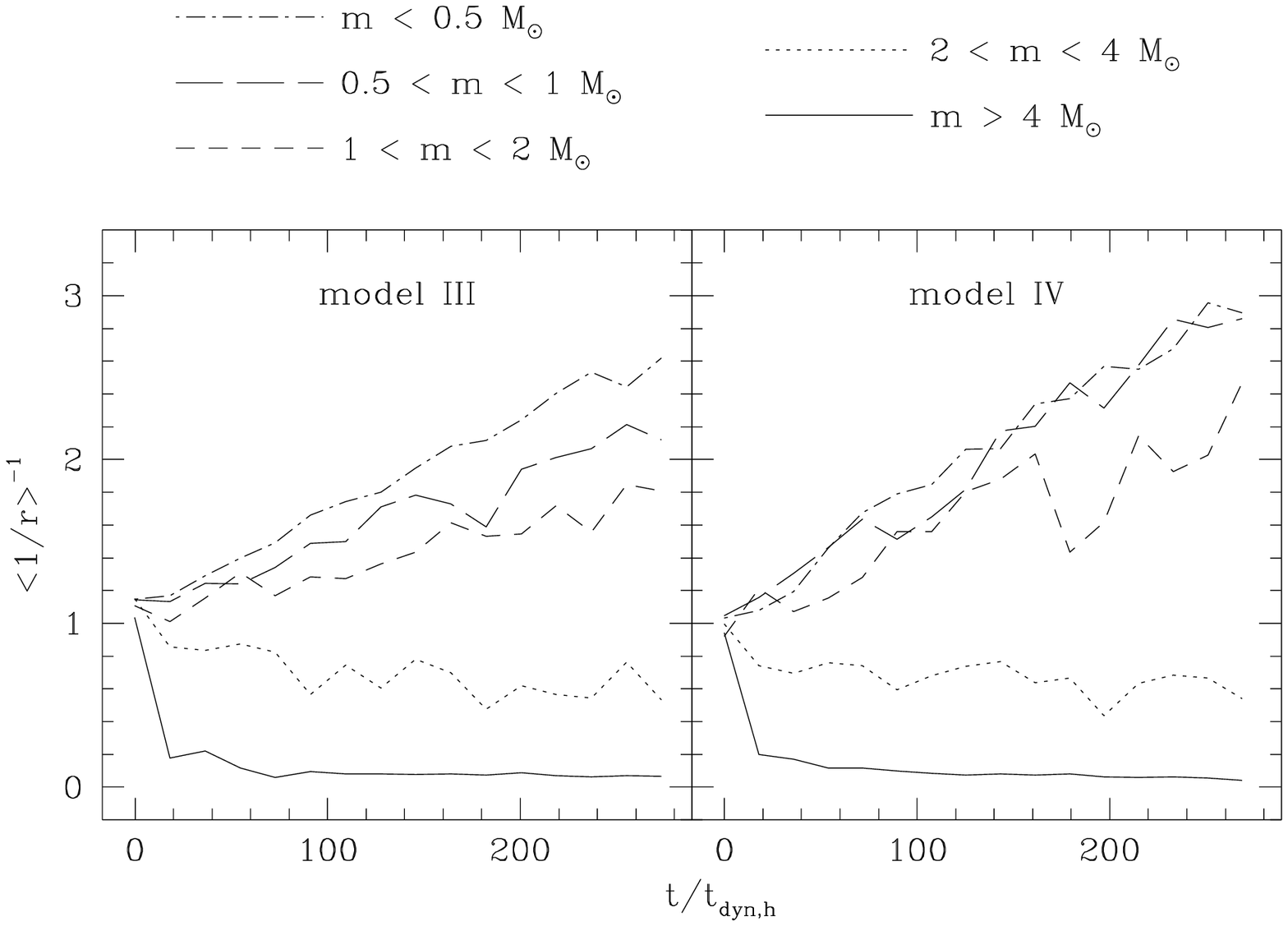}

\figcaption{Characteristic radii of different mass ranges for the multi-mass
models. \label{massdist}}

\clearpage
\plotone{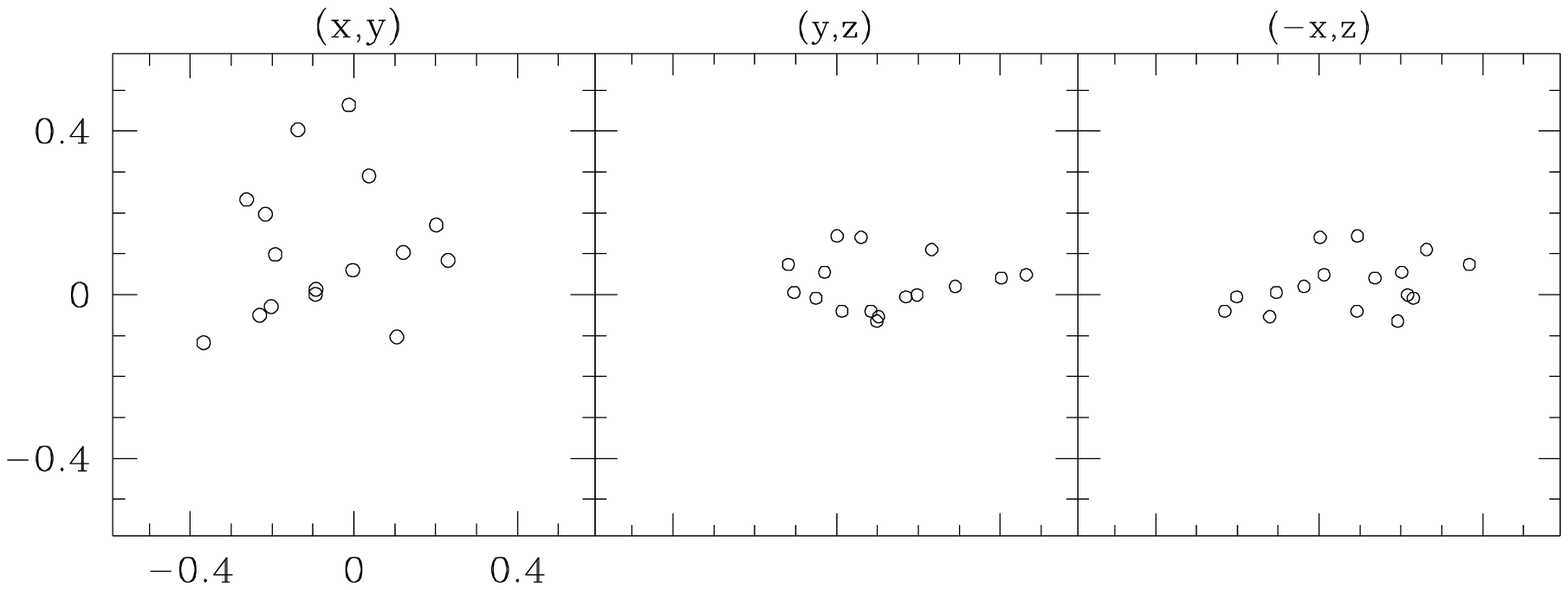}

\figcaption{Lump position at different times for model IV. \label{lumpcm}}

\clearpage
\plotone{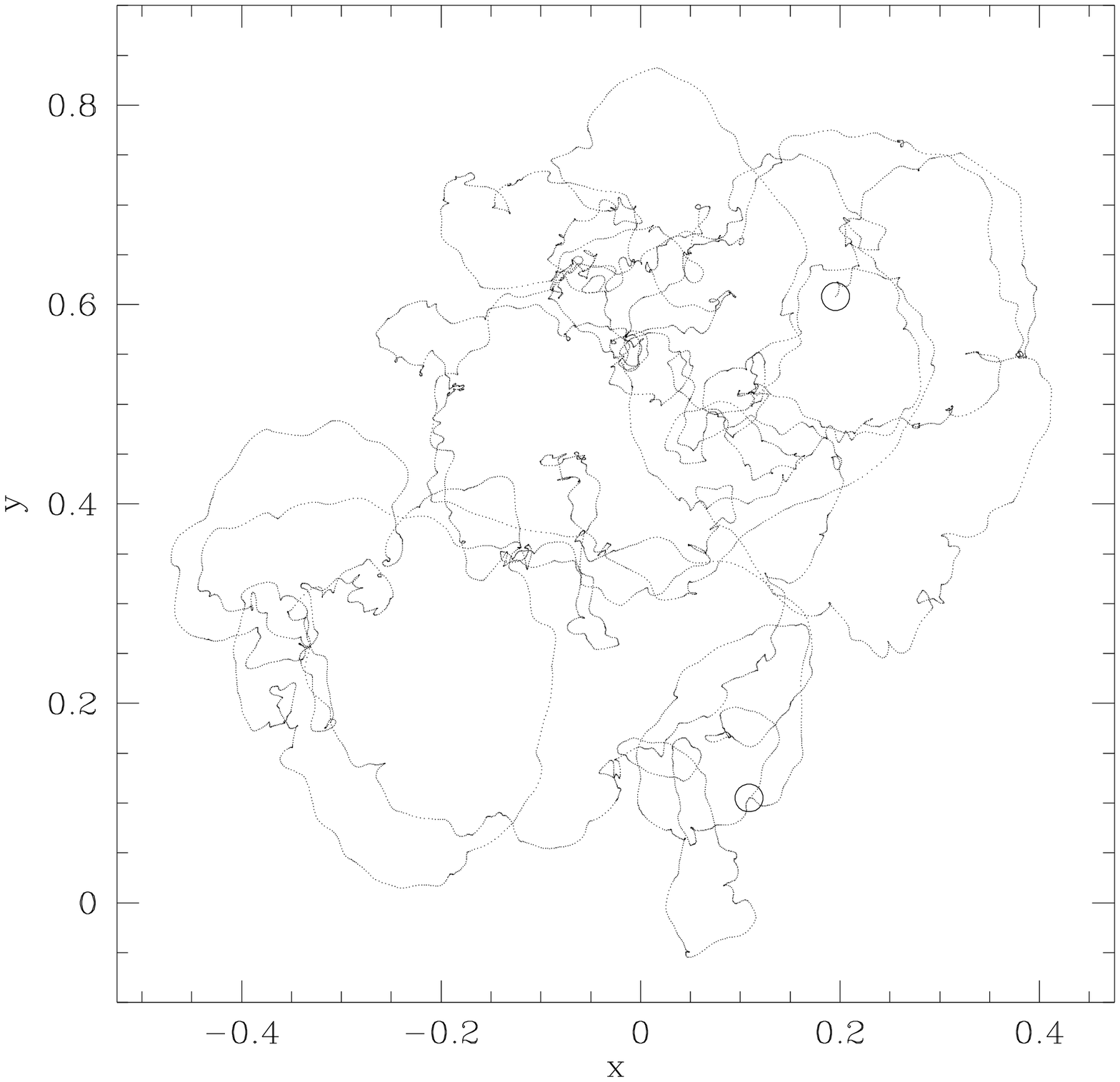}

\figcaption{Position of model IV lump replacement particle. \label{lumprep}}

\clearpage
\plotone{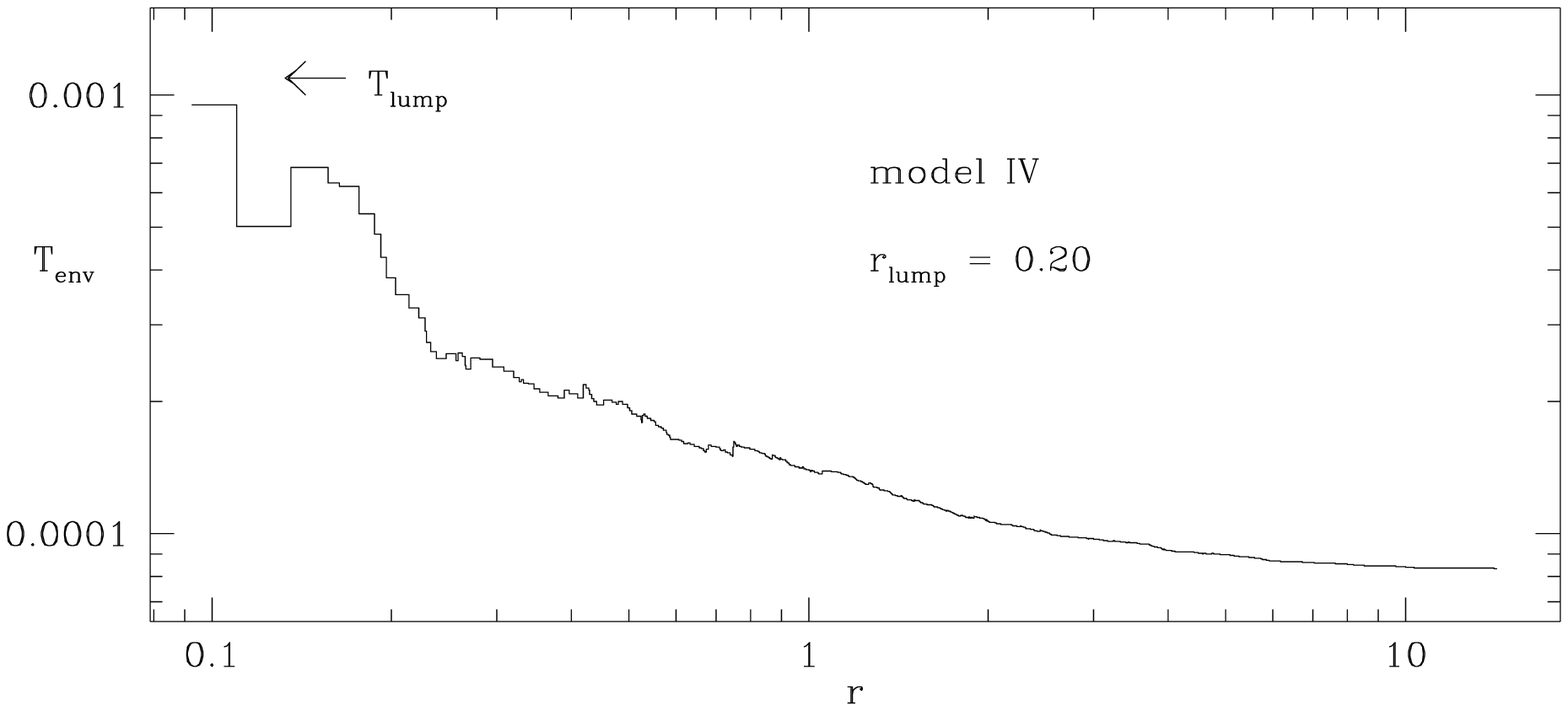}

\figcaption{Lump environment temperature for model IV. \label{lumpT}}

\clearpage
\plotone{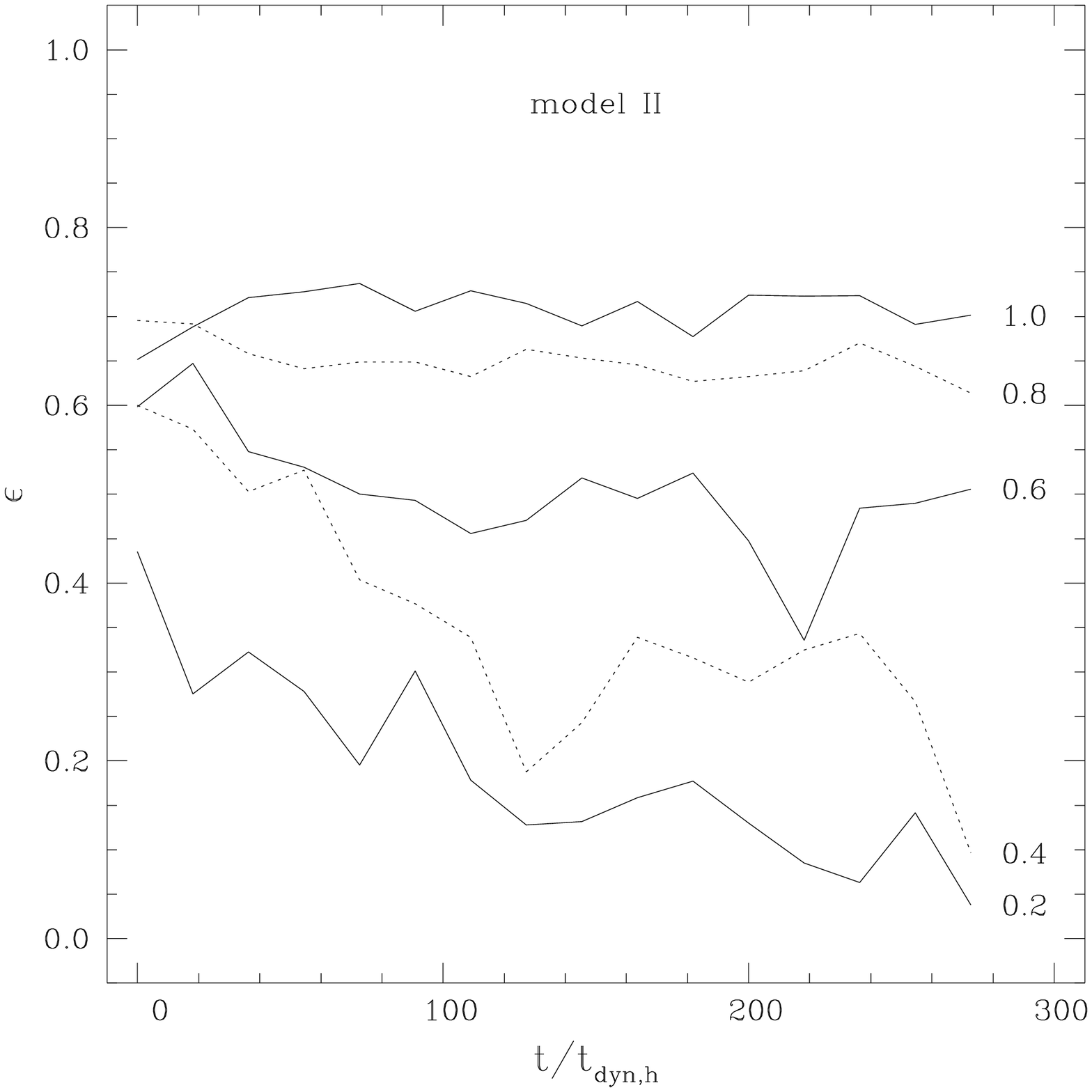}
\figcaption{Flattening versus time for five Lagrangian quintiles of model II
(equal-mass stars, $c/a=0.3$). \label{ellipII}}

\clearpage
\plotone{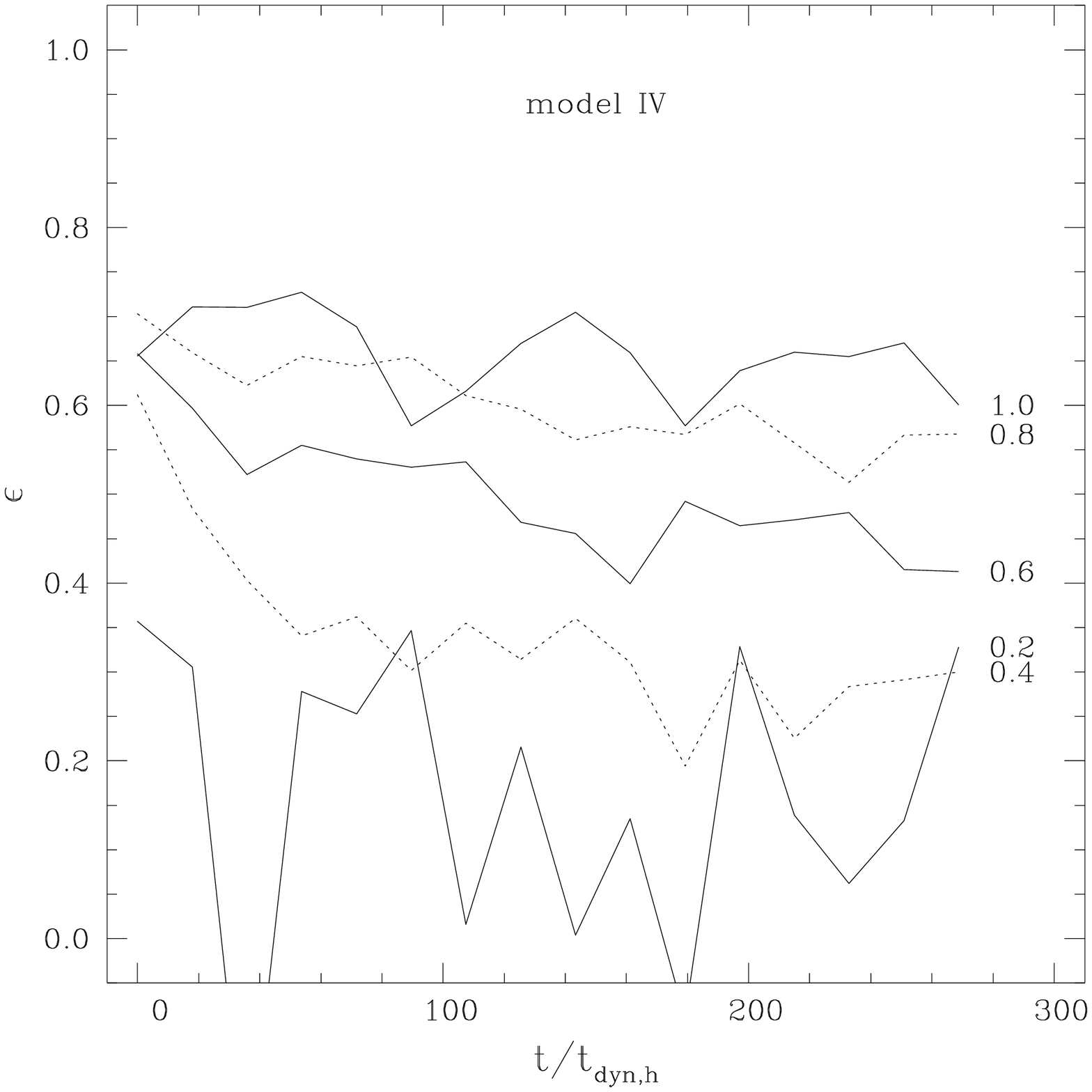}
\figcaption{Flattening versus time for five Lagrangian quintiles of model IV
(Salpeter IMF, $c/a=0.3$). \label{ellipIV}}

\clearpage
\plotone{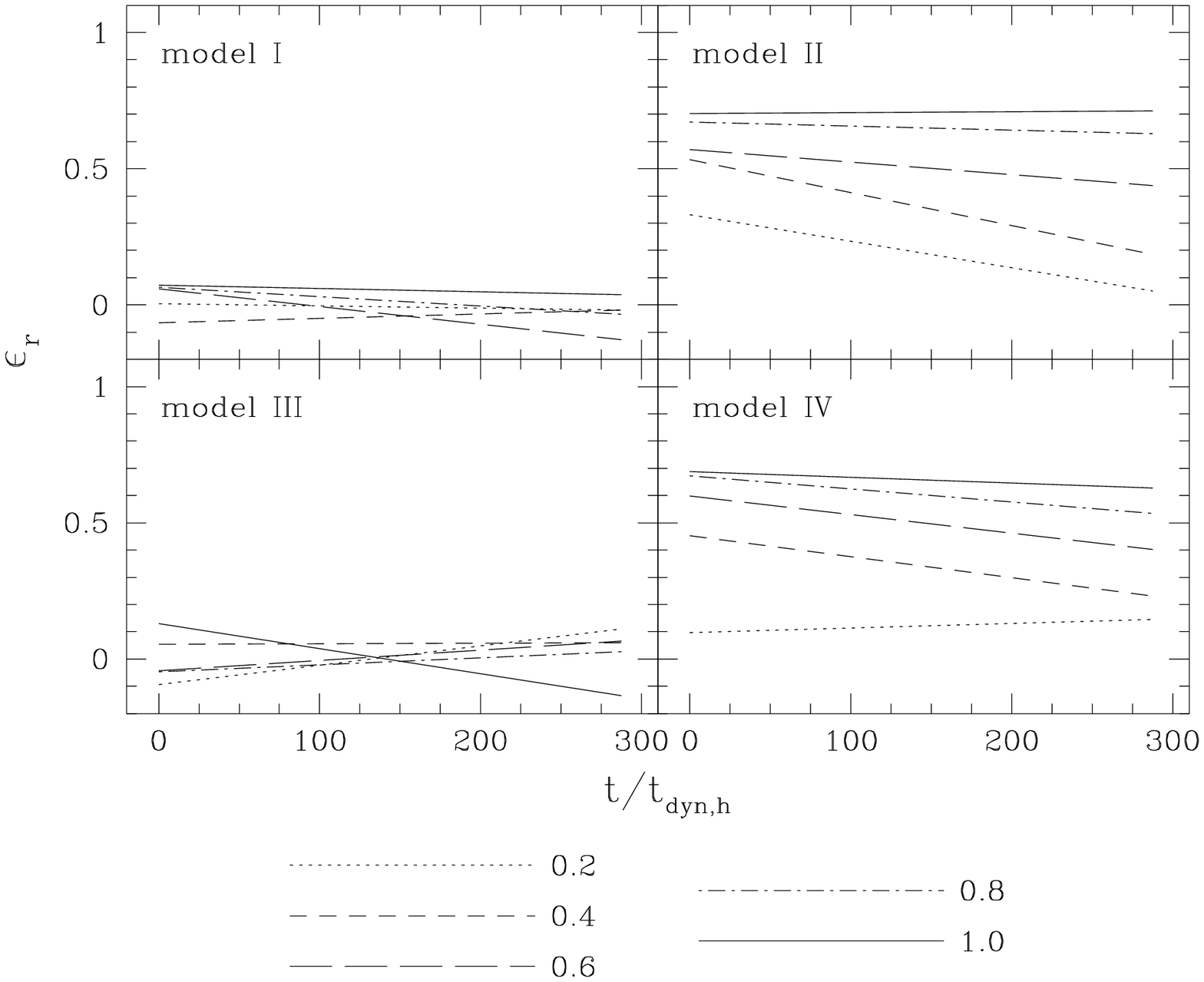}

\figcaption{Linear least-squares fits to the ellipticities of the four
simulations. \label{ellipt}}

\clearpage
\plotone{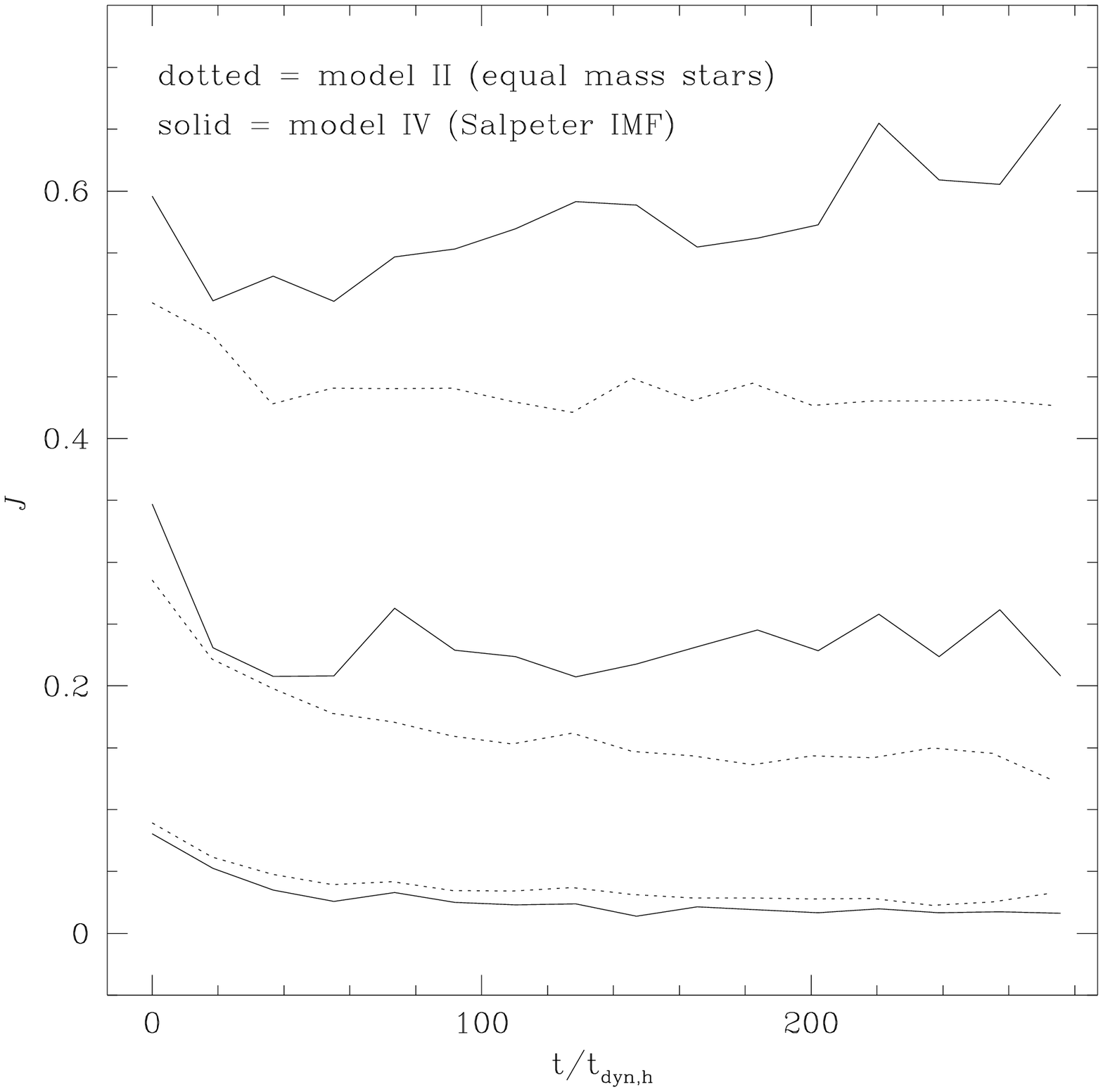}

\figcaption{Angular momentum per unit mass for the inner 3 Lagrangian
quintiles of the rotating systems. \label{J}}

\clearpage
\plotone{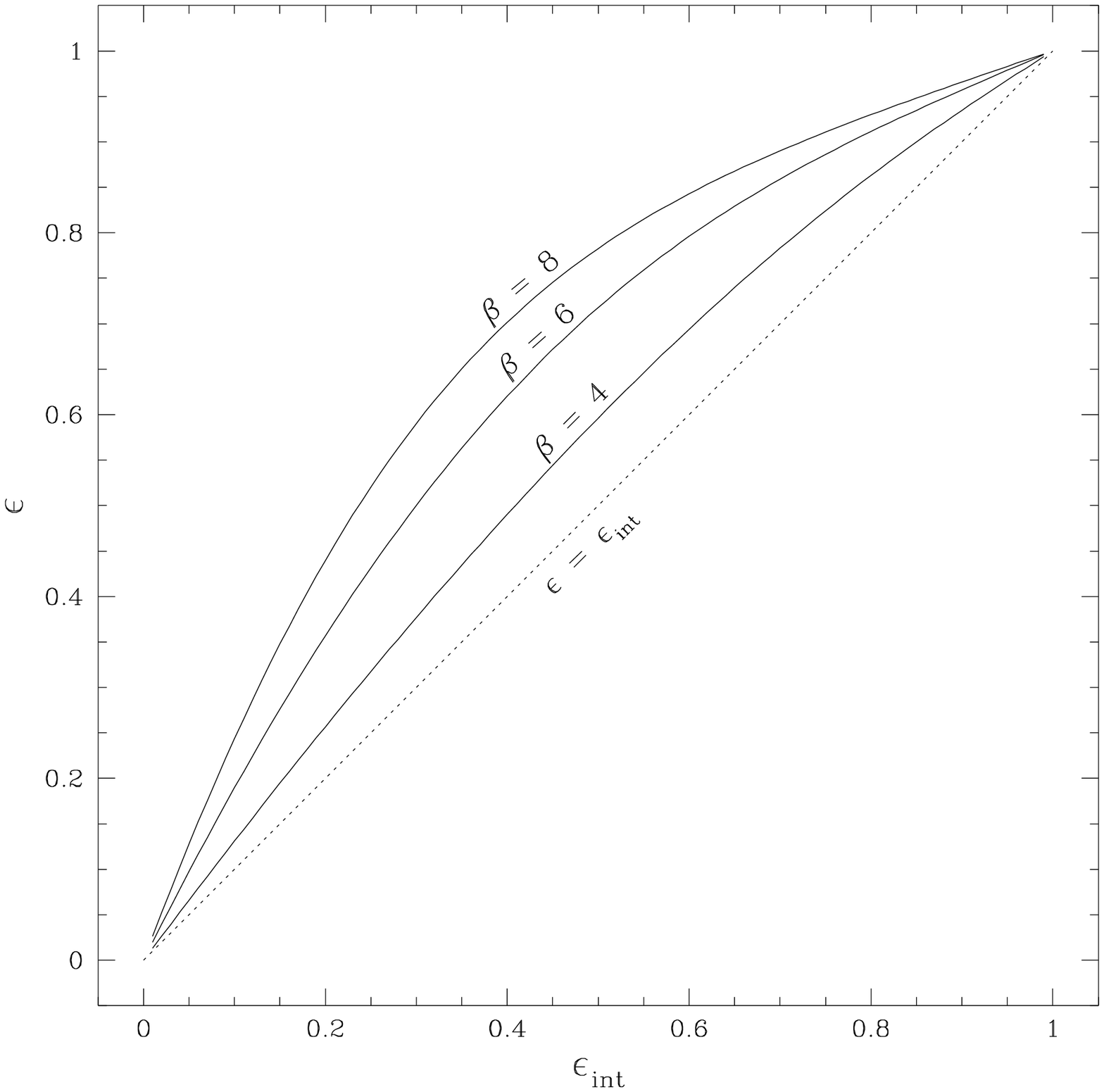}

\figcaption{Calculated thin-shell flatness versus intrinsic ellipticity for
three different power law density profiles. \label{flat_anal}}

\clearpage
\plotone{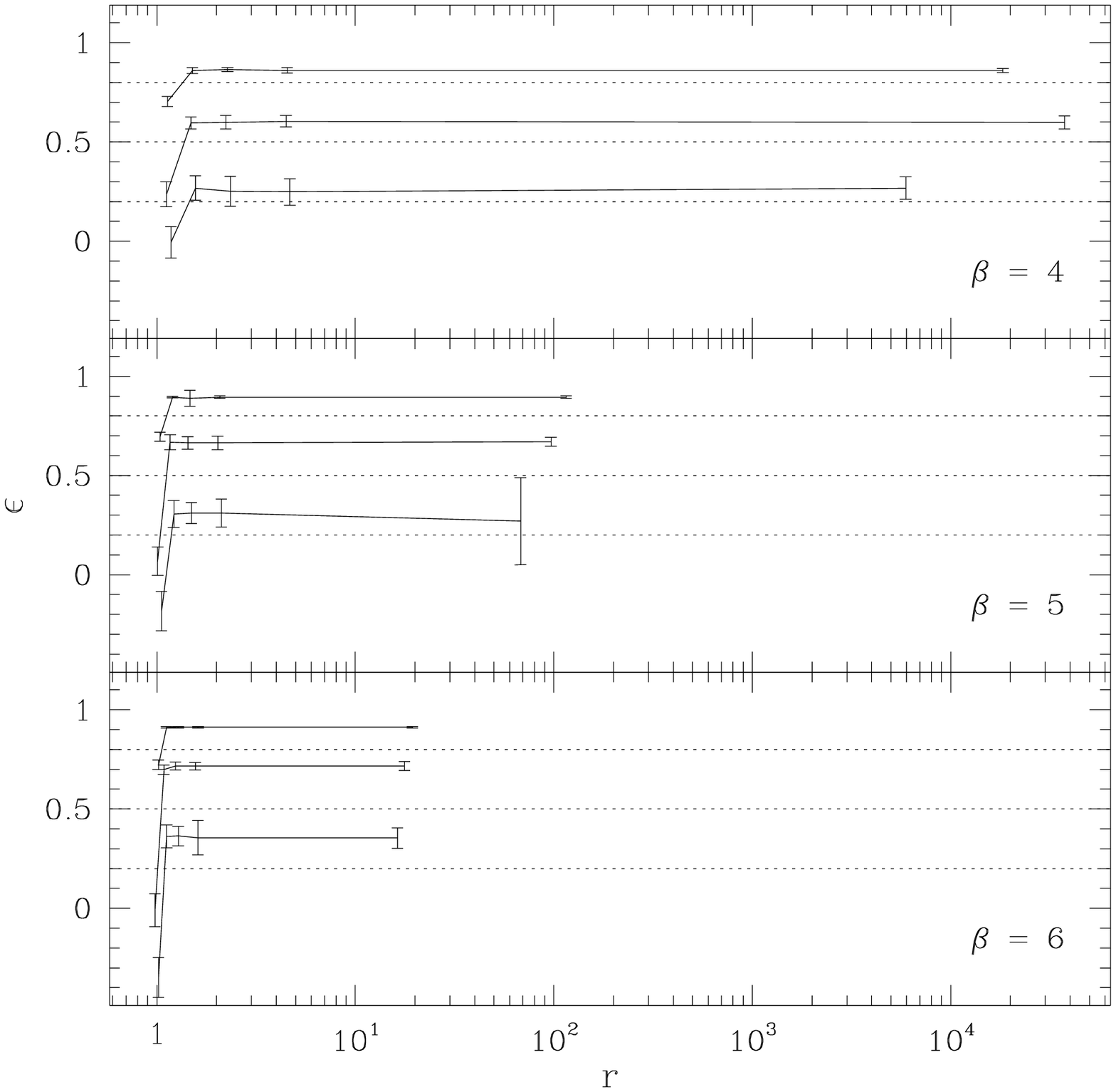}

\figcaption{Calculated thick-shell flatness as a function of shell radius for
$\beta={4,5,6}$.  Within each $\beta$ plot, the curves shown with
one-$\sigma$ error bars are for, from top to bottom,
$\epsilon_{\rm int}={0.8,0.5,0.2}$.  The intrinsic flatness is plotted
with dashed lines. \label{flat_mc}}

\end{document}